\documentclass[%
 aip,
 amsmath,amssymb,
 reprint,%
]{revtex4-1}

\usepackage{graphicx}% Include figure files
\usepackage{subfig}
\usepackage{dcolumn}% Align table columns on decimal point
\usepackage{bm}% bold math
\usepackage[utf8]{inputenc}
\usepackage[T1]{fontenc}
\usepackage{mathptmx}
\usepackage{etoolbox}
\usepackage{amsmath}
\usepackage[ruled,vlined]{algorithm2e}
\usepackage{amsfonts}

\usepackage{siunitx}

\makeatletter
\def\@email#1#2{%
 \endgroup
 \patchcmd{\titleblock@produce}
  {\frontmatter@RRAPformat}
  {\frontmatter@RRAPformat{\produce@RRAP{*#1\href{mailto:#2}{#2}}}\frontmatter@RRAPformat}
  {}{}
}%
\makeatother
\begin{document}

\preprint{AIP/123-QED}

\title[]{Effects of particle-induced electron emission on transport properties in electrically-biased plasma sheaths under fusion-relevant conditions}
% Force line breaks with \\
\author{L. Murillo}
 \email{lucillo@uw.edu, srinbhu@uw.edu.}
 \affiliation{William E. Boeing Department of Aeronautics and Astronautics, University of Washington, Seattle, WA 98195, USA.}%Lines break automatically or can be forced with \\
 \author{C. Skolar}%
\affiliation{Center for Solar-Terrestrial Research, New Jersey Institute of Technology, Newark, New Jersey 07102, USA.}%
\author{K. Bradshaw}%
\affiliation{Department of Astrophysical Sciences, Princeton University, Princeton, New Jersey 08544, USA.}%
\author{B. Srinivasan}%
\affiliation{William E. Boeing Department of Aeronautics and Astronautics, University of Washington, Seattle, WA 98195, USA.}%

% \author{C. Author}
%  \homepage{http://www.Second.institution.edu/~Charlie.Author.}
% \affiliation{%
% Second institution and/or address%\\This line break forced% with \\
% }%

\date{\today}% It is always \today, today,

\begin{abstract}
A rigorous implementation of energy-dependent ion- and electron-induced electron emission in a continuum-kinetic framework is used to reveal their effects on the scaling of plasma properties in the sheath with an applied bias potential to the walls. The approach comes with a novel methodology for modeling particle-induced electron emission (PIEE) that includes 1) improved fitting functions for the yield and spectra, 2) the use of SRIM and a summation of the Lindhard formula and a modified Bethe formula for obtaining accurate stopping powers, and 3) a binding energy correction to the PIEE spectra and ion-induced yield based on density functional theory (DFT) calculations. The emission models are implemented as a fully energy dependent and dynamic boundary condition. For this investigation, tungsten and graphite walls are studied for their relevance in magnetic fusion experiments. Equations are derived from fluid theory that predict the relative importance of ion- and electron-induced emission on the structure of the sheath. The simulations provide evidence for the theoretical predictions, showing that a transition from a classical to space-charge limited (SCL) sheath depends primarily on electron-induced emission. Furthermore, claims in previous literature of increased heat and particle loads to the walls due to electron emission are supported, however differing mechanisms for the increase are observed. The increased thermal and particle fluxes due to PIEE are primarily driven by collisional transfer of energy from emitted electrons in the presheath. Finally, quantitative predictions for the device modeled in this study coincide with previous modeling efforts and experimental measurements.
\end{abstract}

\maketitle

\section{\label{sec:level1}INTRODUCTION}

The study of plasma-material interactions (PMI) in magnetic fusion devices aims to elucidate the complex interplay between thermonuclear plasma exhaust and the solid materials surrounding the plasma. The bombardment of the walls from plasma ions and electrons leads to enormous heat and particle deposition that limits the operational lifetime of the walls while also reducing the performance of the fusion plasma. A specific subset of PMI, electron emission, has been studied extensively in low temperature plasmas, \cite{sydorenko_plasma-sheath_2008,campanell_general_2012,campanell_two_2025,tavant_effects_2018} and some recent work has explored fusion-relevant regimes. \cite{tolias_origin_2020,masline_simulations_2020,moritz_thermionic_2023, bradshaw_effects_2025, skolar_general_2025} Much of the current literature focuses exclusively on thermionic emission, however, this is not the only relevant electron emission channel in fusion devices. Recent advances in kinetic plasma sheath modeling has made it possible to study energy-dependent particle-induced electron emission (PIEE) in great detail. \cite{bradshaw_energy-dependent_2024,bradshaw_effects_2025,skolar_general_2025}

Many open questions remain in terms of how PIEE affects plasma properties close to the walls in fusion reactors. For example, investigation into the influence of PIEE on the sheath potential has been explored with simplified emissive sheath models by Schupfer et al., \cite{schupfer_effect_2006} but the relative influence of ion-induced electron emission (IIEE) on the structure of the sheath compared to secondary electron emission (SEE) in regimes where both types of emission are relevant is not known. Furthermore, it is understood from fluid theory that as PIEE increases the heat and particle fluxes into the wall are expected to increase, \cite{stangeby_plasma_2000,hobbs_heat_1967} however validation of this by a rigorous kinetic treatment of the plasma sheath, particularly in fusion plasmas, is lacking in the literature.

This paper presents a continuum-kinetic model for the plasma sheath that forms near electron emitting surfaces which is leveraged to help answer these questions. The model developed and used in this work is generally applicable across a wide regime of emitting plasma sheaths with and without bias potentials. To understand the implications of emitting sheaths for fusion devices, the parameters relevant to the sheared-flow Z-pinch experiment (FuZE) \cite{zhang_sustained_2019} are used here. The electrodes in a Z-pinch experience large particle fluxes due to the driven plasma current and lack of confinement along the current flow direction, making it an ideal testbed for investigating the physics of emissive sheaths. Moreover, the applied voltage across the plasma leads to large ion impact energies which puts the system in a regime where IIEE becomes relevant. \cite{skolar_general_2025} 

Here analytical and semi-empirical models for IIEE and SEE inform emission boundary conditions which self-consistently calculate the distribution function of emitted electrons. The emitted electron distribution is coupled to a background electron population as a sheath naturally develops. After an initial transient period, state state transport properties are analyzed. These simulations offer new insights into the mechanism behind increasing heat flux with emission in high-pressure discharges, explain discrepancies between theory and experimental measurements of Z-pinch current, and demonstrate inherent differences in how IIEE and SEE couple to the sheath potential gradient. Comparison to previous modeling efforts and experimental measurements of the FuZE device are made, for which the model is found to provide general agreement.

The equations describing how the properties of interest vary with PIEE are derived in Sec. \ref{sec:SheathTheory}. Next, the emission models and their implementation as a boundary condition are discussed in Sec. \ref{sec:emission_models} and \ref{sec:emission_bc}, respectively. The setup, results, and discussion of the emissive sheath simulations are presented in Sec. \ref{sec:setup} and \ref{sec:results}. Finally, concluding remarks regarding the presented findings are made along with directions for future work in Sec. \ref{sec:concl}.

\section{Background}

\subsection{Emissive sheath theory}\label{sec:SheathTheory}

The plasma sheath plays a crucial role in regulating particle transport towards a solid boundary. When plasma strikes the wall, it results in energy loss through thermal conduction and convection, as well as the ejection of impurities from the wall back into the plasma. The regulation of particle transport is dictated by the electric field, or gradient of the electric potential, that is setup due to an initial imbalance of ion and electron fluxes to the wall. The potential gradient within the sheath slows electrons traveling toward the wall, effectively shielding the surface from the full heat load of the core plasma. Understanding how electron emission influences this potential gradient is important, as it can have significant effects on macroscopic transport properties, such as heat and particle fluxes. In this section, the theory describing an emissive sheath in a plasma discharge is developed. The model incorporates both IIEE and SEE for a generalized plasma confined by solid electrodes under varying applied bias potentials.

First, the zero-bias case is examined. Let the current densities for ions and electrons impacting the wall be $J_{i}^{IN}$ and $J_{e}^{IN}$, respectively. If the current density coming out of the wall is solely due to emitted electrons, then it can written as $J_{e}^{EM} = \delta J_{e}^{IN} + \gamma J_{i}^{IN}$, with $\delta$ and $\gamma$ being the ratio of electron current out of the wall to electron and ion currents into of the wall, respectively. These flux ratios are typically referred to as the electron- and ion-induced emission yields, respectively.

From Ref. \onlinecite{stangeby_plasma_2000}, the particle fluxes to the wall are calculated assuming isothermal, Maxwellian electrons, resulting in the following,

\begin{equation}\label{eq:elc_flux}
    \begin{split}
    \Gamma_{e,c}^{IN} & = \frac{1}{4} n_{SE} \bar{c}_e \exp{\left[ \frac{e \Delta\phi_c}{T_e} \right]}, \\ \\
    \Gamma_{e,a}^{IN} & = \frac{1}{4} n_{SE} \bar{c}_e \exp{\left[ \frac{e \Delta\phi_a}{T_e} \right]},
    \end{split}
\end{equation}
where the subscripts $c$ and $a$ here and hereafter denote quantities at the cathode and anode, respectively, $e$ is the elementary charge, $T_e$ is the electron temperature, $n_{SE}$ is the density at the sheath edge, $\bar{c}_e$ is the average electron speed, $\Delta \phi_c$ and $\Delta \phi_a$ are the potential differences across the sheath at the cathode and anode, and $\Delta \phi_c - \Delta \phi_a = \phi_b$ defines the applied bias potential to the walls. Note that temperatures are given in energetic units. For a steady-state sheath, the continuity equation and Bohm criterion prescribe the ion flux,

\begin{equation}\label{eq:ion_flux}
    \Gamma_{i,c}^{IN} = \Gamma_{i,a}^{IN} = n_{SE} c_s.
\end{equation}

Here $c_s$ denotes the ion sound speed. The net current density of electrons $J_{e}^{NET} = J_{e}^{IN} - J_{e}^{EM}$ is given by the following,

\begin{equation}\label{eq:Jnet}
J_{e}^{NET} = e n_{SE} \left[ \frac{(1-\delta) \bar{c_e}}{4} \exp{\left(- \widetilde{\phi}_{p_0} \right)} - \gamma c_s \right],
\end{equation}
where $\widetilde{\phi}_{p_0}$ is the floating plasma potential. The floating wall condition requires the net electron current into the wall to be equal to the ion current into the wall, $J_{i}^{IN} = J_{e}^{NET}$, which can then be solved for the floating or zero-bias plasma potential,

\begin{equation}\label{eq:plasma_potential_NoBias_Emit}
\widetilde{\phi}_{p_0} = - \frac{1}{2} \ln{\left[ \left( 2 \pi \frac{m_e}{m_i} \right) \left( 1 + \frac{T_i}{T_e} \right) \left( \frac{1 + \gamma_0}{1 - \delta_0} \right)^2 \right]}.
\end{equation}

It is convenient to define the zero-bias effective yield $\zeta_0 = (1+\gamma_0)/(1-\delta_0)$ to show how the potential varies with emission when IIEE and SEE are considered. In the more general case, where there can be an arbitrary applied voltage to the walls, there will be some net current in the domain, meaning $J_{i}^{IN} \neq J_{e}^{NET}$ at either wall and in general $\delta_a \neq \delta_c, \gamma_a \neq \gamma_c$. To obtain an expression for the plasma potential in this case the conservation of charge, $J_{i,a}^{IN} + J_{i,c}^{IN} = J_{e,a}^{NET} + J_{e,c}^{NET}$, is utilized,

\begin{equation}
    \begin{split}
    2 e n_{SE} c_s = & e n_{SE} \left[ \frac{(1-\delta_c) \bar{c_e}}{4} \exp{\left(- \widetilde{\phi}_{p} \right)} - \gamma_c c_s \right] \\
    & +  e n_{SE} \left[ \frac{(1-\delta_a) \bar{c_e}}{4} \exp{\left(\widetilde{\phi}_{b} - \widetilde{\phi}_{p} \right)} - \gamma_a c_s \right],
    \end{split}
\end{equation}
leading to the following expression for the plasma potential,

\begin{equation}\label{eq:plasma_potential_Emit}
    \widetilde{\phi}_{p} = - \ln{\left[ \frac{ \zeta^*}{\zeta_0} \exp{\left(- \widetilde{\phi}_{p_0}\right)} \right]},
\end{equation}
where the general effective yield is defined as,

\begin{equation}
    \zeta^* \equiv \frac{2 + \gamma_c + \gamma_a}{1-\delta_c + (1-\delta_a)\exp \left(\widetilde{\phi}_b\right)}.
\end{equation}

The general effective yield $\zeta^*$ depends on the normalized applied bias $\widetilde{\phi}_b = e \phi_b/T_e$ which acts as a scaling factor to the anode SEE term. The decrease in $\widetilde{\phi}_p$ through it's linear dependence on $\delta_a$ is outweighed by the increase due to it's exponential dependence on $\widetilde{\phi}_b$, indicating that the bias potential has a much stronger influence on the plasma potential than the yield at the anode. These competing terms also point to an interesting effect in which space-charge interactions from electrons emitted by the anode are suppressed by an applied bias. This is not a complete picture however, since the particle-induced yield of emitted electrons at either electrode may depend on the bias potential itself. The exact dependence of $\delta_a$ on $\widetilde{\phi}_b$ is not known a priori, but regardless Eq. \ref{eq:plasma_potential_Emit} suggests that anode emission contributes differently to the plasma potential, and likely to a lesser extent, than cathode emission. 

A relatively simple relationship between the one-way heat flux of a half-Maxwellian electron population and the particle flux at the wall is given in Ref. [\onlinecite{stangeby_plasma_2000}, \onlinecite{hobbs_heat_1967}],

\begin{equation}\label{eq:elc_heat_flux}
    \begin{split}
    q_{e,c} & = \eta_c \Gamma_{e,c} T_{e}, \\ \\
    q_{e,a} & = \eta_a \Gamma_{e,a} T_{e},
    \end{split}
\end{equation}
where $q_e$ and $q_i$ are the electron and ion heat flux, respectively, the heat transmission coefficient, typically denoted $
\gamma$ but to avoid confusion $\eta$ is used here, is defined as $\eta_c = 2 \zeta_c (1 - \delta_{e,c}) + |\widetilde{\phi}_p|$ at the cathode and $\eta_a = 2 \zeta_a (1 - \delta_{e,a}) + |\widetilde{\phi}_p - \widetilde{\phi}_b|$ at the anode. The effective yield at the anode and cathode used here are defined as $\zeta_{c,a} = (1+\gamma_{c,a})/(1-\delta_{c,a} + \delta_{e,c,a})$. A distinction is made between the secondary electrons captured by $\zeta_{c,a}$, and the backscattered electrons captured by $\delta_e$, both of which carry heat away from the walls, because the secondary electrons are typically lower energy.

For the sheared-flow Z-pinch, a strong scaling between the production of fusion neutrons and the current flowing through the plasma has been demonstrated. \cite{shumlak_z-pinch_2020} Following the theory presented thus far, the net plasma current, assuming the plasma takes the form of an axisymmetric pinch, is expressed by,

\begin{equation}\label{eq:pinch_current}
    \begin{split}
    I_p & = \pi r_p^2 e n_{SE} \left[ (1 + \gamma_c) c_s - (1 - \delta_c) \frac{1}{4} \bar{c}_e \exp{\left( - \widetilde{\phi}_{p} \right)} \right] \\ \\
    & = \pi r_p^2 e n_{SE} \left[ (1 - \delta_a) \frac{1}{4} \bar{c}_e \exp{\left( \widetilde{\phi}_b - \widetilde{\phi}_{p} \right)} - (1 + \gamma_a) c_s \right],
    \end{split}
\end{equation}
where $r_p$ is the radius of the pinch. In general, the $\pi r_p^2$ term will be replaced by the cross-sectional area of the plasma perpendicular to the direction of the current density, assuming a constant current density with radial position as is done here. For increasingly large bias potentials, the impacting electron contribution to the current at the cathode in Eq. \ref{eq:pinch_current} vanishes, leaving the saturation current,

\begin{equation}\label{eq:Isat}
    I_{sat} = \pi r_p^2 e n_{SE} c_s (1 + \gamma_c).
\end{equation}

The saturation current is proportional to the IIEE yield at the cathode, however, $\gamma_c$ varies with the applied bias as well. Without knowledge of how the yield depends on the bias potential, the true saturation current can not be predicted from this theory alone. For large but still finite bias potentials, if the IIEE yield is assumed to have a linear relationship with the applied bias then the current ``saturates'' to a line with slope $\pi r_p^2 e n_{SE} c_s [\gamma_c(\widetilde{\phi}_b) - \gamma_0]/\widetilde{\phi}_b$. A numerical analysis can be done to verify that the anode saturation current is equivalent to Eq. \ref{eq:Isat}. To find the actual relationship between the PIEE yield and applied bias, a rigorous model of ion- and electron-induced emission is needed.

\subsection{\label{sec:emission_models}Ion- and electron-induced emission}

When ions or electrons impinge on a solid material, there is a probability that some number of electrons will be emitted from the surface. These emitted electrons ultimately arise from discrete interactions between the projectile and the material's atomic lattice. Screened collisions of the projectile with target atoms results in excitations of bound or conduction electrons and recoiling nuclei. As liberated electrons, recoiling nuclei, and the primary projectile diffuse through the material they undergo similar collision events with other target atoms, potentially liberating more electrons and sending more nuclei recoiling in the process. 

This description of the production of free electrons in the material is referred to as an ionization cascade model. A fraction of the liberated electrons produced in a collision cascade will escape the surface. These emissions are referred to as kinetic emissions. Collisions are not the only way for impacting charged particles to produce free electrons in a material. For example, in the Auger neutralization process, electrons can tunnel into the ion's lowest available bound state, releasing energy to a surrounding electron. \cite{baragiola_principles_1982} The free electrons produced in these types of processes that diffuse out of the material are referred to as potential emissions. Potential emission is in general negligible for bounded, high temperature plasmas comprised of singly-charged ions such as hydrogen, but can become appreciable for massive, low energy, multiply-charged ions. \cite{bercx_quantitative_2019,stockl_separation_2004,fehringer_potential_1987} Hydrogen is the only species considered in this study and high energy impacts are expected, therefore only kinetic emissions are modeled.

\subsubsection{Ion-induced emission}

In Jorgen Schou's seminal paper \cite{schou_transport_1980} the yield and energy spectrum of emitted electrons were derived from kinetic transport theory for an arbitrary target material and impacting ion species. The emitted electron energy spectrum is defined as the derivative of the yield $\gamma$ with respect to the emitted electron energy $E_e$ and is given by the following,

\begin{equation}\label{eq:dgammadE}
    \begin{split}
    \frac{d\gamma}{dE_e} & = \frac{\Gamma_m E_e
    S_{i,e}(E_i)}{4 (E_e + W)^2 S_{e,e}(E_e + W)}, \\ \\
    & \Gamma_m = \frac{m}{\psi(1) - \psi(1-m)},
    \end{split}
\end{equation}
where $\psi(x)$ is the digamma function, $W$ is the energy barrier height, $m$ is related to the energy at which the spectrum peaks by $m = 2 - 0.5(W/E_{peak})$, $E_i$ is the ion impact energy, and $S_{i,e}$ and $S_{e,e}$ are the electronic, or inelastic, stopping power for ions and electrons, respectively. Integrating the spectrum then provides the IIEE yield,

\begin{equation}\label{eq:gamma}
\gamma  = \int_0^\infty{\frac{\Gamma_m E_e
 S_{i,e}(E_i)}{4 (E_e + W)^2 S_{e,e}(E_e + W)} \ dE_e}.
\end{equation}

The energy barrier height is defined as the energy necessary for an electron in the material to escape the surface. For metals this is taken to be the sum of the work function $\Phi_W$ and fermi energy $E_F$. The stopping powers, defined as the average energy lost by the impacting charged particle per unit distance traveled, describe how the ions and liberated electrons transfer energy to target atoms. Analytical models for stopping power were derived from first principles many decades ago. \cite{bohr_ii_1913,bethe_zur_1930,lindhard_stopping_1964} 

A commonly used model for the stopping power is the Bethe-Bloch formula along with its many corrections. \cite{bloch_zur_1933} This formula was originally derived from quantum theory with the first Born approximation, assuming a weak scattering potential. At lower impact energies, the approximation fails and elastic scattering becomes dominant in slowing the projectile down. To remedy this, some corrections were added to take into account material as well as relativistic effects, resulting in the following, \cite{salvat_bethe_2022}

\begin{equation}\label{eq:BB_corrected}
S_{i,e}^{BB} = \frac{4 \pi e^4 Z N}{m_e v^2} \ln{ \left( \frac{2 m_e v^2 \gamma_{rel}^2}{I} - \beta^2 + L_2^{NR} - \frac{C}{Z} - \frac{\delta_F}{2} \right) }
\end{equation}
where $m_e$ is the electron mass, $N$ is the material's atomic number density, $v$ is the projectile velocity, $\gamma_{rel}$ is the Lorentz factor, $\beta=v/c$, $Z$ is the material's nuclear charge state, $I$ is the mean excitation energy, and $L_2^{NR}$, $C/Z$, and $\delta_F/2$ are the Bloch, shell, and density effect corrections, respectively. Note that this form assumes a singly charged impacting species. These correction terms are difficult to calculate and often require experimental data to inform them for a given material. \cite{salvat_bethe_2022}

The corrections improve the accuracy down to impacting velocities comparable to bound electron orbital velocities, however for lower impacting energies the formula still diverges from experimental measurements. In 1964, Jens Lindhard published a semi-classical derivation of the stopping power which exhibits better accuracy at low impact energies. \cite{lindhard_stopping_1964} Typically, when the stopping power over a wide range of energies is needed, some clever stitching of the two formulas is performed. Here it will be shown that a simple summation of the Lindhard and Bethe-Bloch formulas provides good agreement with experimental data when calculating the yield and spectra using Eqs. \ref{eq:dgammadE} and \ref{eq:gamma}, similar to what was proposed in Ref. \onlinecite{Haque2019}. The Lindhard formula is expressed as,

\begin{equation}\label{eq:Lindhard_dEdx}
S_{i,e}^{L} =  \frac{4 \pi e^4 n}{m_e v^2} L
\end{equation}
where $n$ is the free electron number density and $L$ is the dimensionless stopping number given by,

\begin{equation}\label{eq:Lindhard_L}
    \begin{split}
  L & = \frac{i}{\pi \omega_p^2} \int_0^\infty{\frac{1}{k} \ dk} \int_{-kv}^{kv}{\omega \ d\omega} \ \left( \frac{1}{\epsilon(k,\omega) - 1} \right) \\ \\
    & = \frac{6}{\pi} \int_0^\frac{v}{v_F}{u \ du} \int_0^\infty{dz \ \frac{z^3 f_2(u,z)}{\left[ (z^2 + \chi^2 f_1(z,u))^2 + (\chi^2 f_2(z,u))^2 \right]^2}}.
    \end{split}
\end{equation}

The plasma frequency $\omega_p = \sqrt{4 \pi n e^2 / m_e}$ in the denominator indicates that this model includes collective effects from the material. In the second form of $L$, the wavenumber and frequency are converted to non-dimensional variables $z = k / 2k_F$ and $u = \omega / k v_F$, where $k_F$ and $v_F$ are the Fermi wavenumber and velocity, respectively. The term $\chi$ is related to the Fermi velocity by $\chi^2 = e^2 / \pi \hbar v_F$. Finally, the term including the dielectric function $\epsilon(k,\omega)$ is rewritten in terms of $z$ and $u$ using the functions $f_1$ and $f_2$, which were derived from a first-order perturbative quantum mechanical treatment of the system,

\begin{equation}\label{eq:Lindhard_f1}
    \begin{split}
    f_1(z,u) = \frac{1}{2} & + \frac{1}{8z} \left[ 1 - (z-u)^2 \right] \ln{\left| \frac{z-u+1}{z-u-1} \right|} \\ & + \frac{1}{8z} \left[ 1 - (z+u)^2 \right] \ln{\left| \frac{z+u+1}{z+u-1} \right|},
    \end{split}
\end{equation}

\begin{equation}\label{eq:Lindhard_f2}
    f_2(z,u) =
    \begin{cases}
        \frac{\pi}{2} u & \text{for } z+u < 1\\
        \frac{\pi}{8z} \left[ 1 - (z-u)^2 \right] & \text{for } |z-u| < 1 < z+u\\
        0 & \text{for } |z-u| > 1
    \end{cases}
    \ .
\end{equation}

Here the materials of interest are graphite and tungsten, which are used for the electrodes in FuZE and are ubiquitous in fusion experiments. The free electron model is used to obtain the Fermi speed for graphite, \cite{kittel_introduction_nodate}

\begin{equation}\label{eq:Fermi_Energy}
v_F^2 = \frac{\hbar^2}{m_e^2} (3 \pi^2 n)^{\frac{2}{3}},
\end{equation}
where $n = N$ is found to have good agreement with experiment. \cite{orlita_magneto-transmission_2008} For tungsten, the Fermi speed can be found using the Fermi energy, \cite{rai_theoretical_2014} $E_F = \frac{1}{2} m_e v_F^2 = $ \qty{9.75}{\electronvolt}.

The stopping number is found by numerical integration of Eq. \ref{eq:Lindhard_L}, the details of which are discussed in Sec. \ref{sec:emission_bc}. Due to the complexity of obtaining the corrections to the Bethe-Bloch formula, the high-energy stopping power for the ions and low-energy stopping power of the cascade electrons are calculated using the modified Bethe formula from Nguyen-Truong, \cite{nguyen-truong_modified_2015}

\begin{equation}\label{eq:bethe_nguyen_correction_SI}
S_{e,e}^{NT} = \frac{2 \pi Z e^4 N }{ (4 \pi \epsilon_0)^2 E} 
\ln \bigg[  \sqrt{\frac{\mathtt{e}}{2}} \frac{E}{I}  + G(E)\bigg], 
\end{equation}

% \begin{widetext}
\begin{equation}\label{eq:G_nguyen}
    \begin{split}
    & G(E) =  \ 1 - \sqrt{\frac{\mathtt{e}}{2}} \ln \bigg[ 1 + \Big( \frac{E}{I} \Big)^2 \bigg] \frac{I}{E} \\
    & + \frac{1}{3} \ln \Big( \frac{Z}{2} \Big)
    \exp \bigg[ - \frac{3}{\sqrt{Z}} \Big( 1 - \frac{2}{\sqrt{Z}} + \ln \frac{E}{I} \Big)^2  \bigg] \frac{E}{I}.
    \end{split}
\end{equation}
% \end{widetext}

The electronic stopping power for electrons $S_{e,e}^{NT}$, given in SI units, depends on the impinging electron energy $E$, mean excitation energy $I$, target material atomic number $Z$ and number density $N$. Here $\epsilon_0$ is the vacuum permittivity and $\textbf{e}$ is Euler's number. The formula can be adapted for ion stopping if the factor of $2 \pi$ is changed to $4 \pi$ and the electron energy $E$ is replaced with $m_e v_i^2$, where $v_i$ is the impacting ion velocity. \cite{carron_introduction_2006} This is found to agree well with Eq. \ref{eq:BB_corrected}.

A reliable tool for calculating stopping powers for various ions and target materials is the Monte Carlo code SRIM, \cite{ziegler_srim_2015} which uses a density averaged Lindhard formula at lower energies and the corrected Bethe-Bloch formula, shown in Eq. \ref{eq:BB_corrected}, at higher energies. The Lindhard-Bethe sum (LBS) is found to follow closely with results from SRIM, and a comparison of the two approaches against experimental measurements is provided in the Sec. \ref{sec:emission_bc}.

Up until now, the transport physics of recoiling target electrons and nuclei has been ignored. In the general form of Eq. \ref{eq:dgammadE}, the $S_{i,e}$ term becomes linear combination of the electronic and nuclear components of the stopping power called the spatial energy distribution (evaluated at the material's surface),

\begin{equation}\label{eq:D_transport}
D_{surf}(E_i) = \beta_e S_{i,e}(E_i) + \frac{\beta_r \eta_t(T_{max})}{T_{max}} S_{i,n}(E_i).
\end{equation}
where $S_{i,n}$ is the ion nuclear stopping power, $\beta_e$ and $\beta_r$ are electronic and nuclear recoil transport coefficients, $T_{max}$ is the maximum energy transferred in ion-target atom collisions, and $\eta_t$ is the total energy deposited by the projectile into recoiling nuclei. These transport terms take into account the energy lost by the primary to the material that does not contribute to electronic excitation.

For energetic, light ions $D_{surf}$ reduces to the ion's electronic stopping power \cite{holmen_direct_1979} $S_{i,e}$ leading to Eq. \ref{eq:dgammadE}. Neglecting the transport physics provides a simpler approach to calculating $d\gamma/dE_e$ that still exhibits good agreement with experiment as will be demonstrated in subsequent sections. 

Eq. \ref{eq:dgammadE} assumes all liberated electrons originate from the conduction band and therefore does not take into account electron binding energy to nuclei of target atoms. For non-metals, this is not an accurate description as the effective energy barrier can be severely under-predicted and the yield over-predicted. In Schou's derivation of Eq. \ref{eq:dgammadE} an alternate form with a correction is introduced,

\begin{equation}\label{eq:dgammadE_bind}
\frac{d\gamma}{dE_e}  =\frac{\Gamma_m E_e
 S_{i,e}(E_i)}{4 (E_e + W)(E_e + W + (2-m)|V|) S_{e,e}(E_e + W)},
\end{equation}
where $V$ is the binding energy. However, it is not clear what value for $V$ should be used for a given material. In this work, an average binding energy is used in evaluating IIEE for graphite. This is a reasonable solution as the yield is already calculated from averaged quantities, such as the stopping powers. Semimetals, like graphite, and insulators have low conduction electron densities, meaning a significant fraction of emitted electrons will originate from bound states \cite{fernandez-coppel_fully_2024} and therefore the corrected form of $d\gamma/dE_e$, given by Eq. \ref{eq:dgammadE_bind}, must be used. 

To calculate the average binding energy, the density of states (DOS) for a given material is utilized. Integrating the product of the Fermi-Dirac distribution and the DOS over all binding energies gives the total number of electrons. The product of the DOS and Fermi-Dirac distribution are used as weights in a weighted average to calculate $V$,

\begin{equation}\label{eq:bind_energy}
V = \langle E \rangle = \frac{\int E f_F(E) D(E) \ dE}{\int f_F(E) D(E) \ dE}.
\end{equation}
where $D(E)$ is the DOS and $f_F(E)$ is the Fermi-Dirac distribution given by,

\begin{equation}
f_F(E) = \frac{1}{1 + \exp{\left[ -(E - E_F)/T \right]}}.
\end{equation}

For this work, density functional theory calculations by Klintenburg et al. \cite{klintenberg_evolving_2009} for the density of states in graphite are used to calculate $V = -9.3 \ eV$ relative to the Fermi level. The temperature dependence of $V$ from $f_F$ on $\gamma$ is found to be negligible.

\subsubsection{Secondary electron emission}

The emission of electrons from a material surface by electron impacts has arguably been studied more extensively than for ion impacts. Much of the early literature suggests that the SEE yield should be proportional to the electronic stopping power as is the case for IIEE. \cite{baroody_theory_1950,lye_theory_1957} Furman and Pivi developed a probabilistic model for SEE that takes into account the elastically backscattered and inelastically rediffused electrons when calculating the SEE yield and energy spectrum. \cite{furman_probabilistic_2002} These populations of electrons differ from the so-called true secondaries, which originate from the material, in that they result from the impacting electrons that are deflected away from the surface by elastic or inelastic collisions with target atoms. This model, while phenomenological, does not require the stopping power for a given material, making it much simpler to use if data for the material is available. 

The true secondary population is described by the following semi-empirical fitting function,

\begin{equation}\label{eq:fp_see_yield}
\begin{split}
\delta_{ts} & (E_p, \mu_p) = \hat{\delta}(\mu_p)D(E_p/\hat{E}(\mu_p)), \\
& \hat{\delta}(\mu_p) = \hat{\delta}_{ts}[1 + t_1(1 - \mu_p^{t_2})], \\
& \ \ \hat{E} = \hat{E}_{ts}[1 + t_3(1 - \mu_p^{t_4})], \\
& \quad \ \ D(x) = \frac{sx}{s - 1 + x^s},
\end{split}
\end{equation}
where $\delta_{ts}$ is the true secondary yield and $\hat{\delta}_{ts}$, $\hat{E}_{ts}$, $s$, $t_1$, $t_2$, $t_3$, and $t_4$ are material-dependent fitting parameters. Only normal impacts are considered here, so the direction cosine of an impacting electron is $\mu_p = 1$. For this work, the Farhang et al. \cite{farhang_electron_1993} and Walker et al. \cite{walker_secondary_2008} datasets are used to calculate the SEE fitting parameters for graphite and tungsten, respectively. The fit for the backscattered population is given by,

\begin{equation}\label{eq:delta_e}
    \begin{split}
    \delta_e (E_p) & = P_{1,e}(\infty) \\
    + \left( 
    \widehat{P}_{1,e} - P_{1,e}(\infty) \right) & \exp{\left[ -  \frac{\left( (|E_p - \widehat{E}_e|)/W \right)^p}{p} \right]}.
    \end{split}
\end{equation}
where $\delta_e$ is the backscattered yield and $P_{1,e}$, $P_{1,e}(\infty)$, $\widehat{E}_e$ $W$, and $p$ are the fitting parameters. The fit for the rediffused population is,

\begin{equation}\label{eq:delta_r}
    \delta_r (E_p) = P_{1,r}(\infty)b\left[1 - \exp{(E_0/E_r)^r} \right].
\end{equation}
where $\delta_r$ is the rediffused yield and $P_{1,e}(\infty)$, $E_r$, and $r$ are fitting paramters. Because reflection data used for obtaining these fits often can not discriminate between the backscatterd and rediffused populations, a sum of Eqs. \ref{eq:delta_e} and \ref{eq:delta_r} is fit to the data. Constraints from experimental measurements and numerical modeling can help guide the calculation of fitting parameters in regions of scarce data. For example, the initial guesses for the least-squares calculation of the rediffused fitting parameters for tungsten were informed by Monte Carlo calculations from Yang et al. \cite{yang_electron_2021}

Previous modeling of the emissive sheath with conducting materials \cite{bradshaw_energy-dependent_2024,skolar_general_2025} demonstrate the shape of the SEE spectrum is well described by the Chung-Evarhart distribution, \cite{chung_simple_1974}

\begin{equation}
f_{CE}(E_e) = \frac{E_e}{(E_e + \Phi_W)^4}.
\end{equation}

It is important to note though that if the ion stopping power is divided out of Eq. \ref{eq:dgammadE}, what remains is a material dependent function that determines the shape of the electron-induced spectrum. \cite{schou_secondary_1987} The potential application of the Schou model to the SEE spectrum, comparison to the Chung-Everhart model, as well as how these models are implemented in a numerical scheme will be discussed in Sec. \ref{sec:emission_bc}.

\subsection{\label{sec:level3}Space-charge limited and inverse sheaths}

It has been shown experimentally and can be shown from the theory presented here that most emitted electrons possess a relatively low energy, on the order of \qty{1}{\electronvolt} for most materials. Emitted electrons are gradually accelerated by the sheath electric field. However, given their slow initial velocities relative to the electron thermal speed, they will tend to accumulate near the wall increasing the local electron density. This increase in the electron density changes the potential profile and weakens the electric field via Poisson's equation. The classical sheath is characterized by a monotonically decreasing potential profile. If the yield of emitted electrons becomes significant enough, a fundamental restructuring of the sheath, away from the classical structure, occurs. \cite{hobbs_heat_1967}

One of these alternative sheath structures is the space-charge limited (SCL) sheath. In this case, the emitted electrons drive the electric field to be less than or equal to zero at the wall and the potential profile becomes non-monotonic. \cite{takamura2004} Near the wall a localized potential well, sometimes called a virtual cathode, forms in order to reflect some of the emitted electrons back to the wall in an attempt to equalize the electron and ion fluxes. A consequence of the weakening of the sheath electric field is that lower energy electrons are allowed to stream towards the wall and the particle flux increases. This is taken to the extreme if the sheath becomes inverted. In the inverse sheath, the potential becomes monotonic once more, but now increases towards the wall rather than decreasing. As a result, electrons are now accelerated to the wall and ions are reflected. Campanell and Umansky \cite{campanell_strongly_2016} show that an SCL sheath transitions to an inverse sheath as cold ions become trapped in the virtual cathode. The presented kinetic simulations in this paper do not include any significant source of cold ions in the sheath, from charge exchange reactions or ion backscattering for example, and so no inverse sheath is expected to develop.

From the derived potential, Eq. \ref{eq:plasma_potential_NoBias_Emit}, it is clear that IIEE and SEE contribute differently to changes in sheath structure. A more thorough analysis of why this occurs will be explored in a future paper. Intuitively though, one can imagine that as ions impact the wall and produce emitted electrons that in turn weaken the electric field, then the electron flux to the wall increases in response and the local electron density decreases again as they are absorbed by the wall. In the case where electron-induced emission is instead considered, after the electric field is weakened by emitted electrons, the increase in electron flux to the wall drives further emission and does not allow the system to return to its initial state as easily.

\section{Emission boundary conditions}\label{sec:emission_bc}

Modeling of the plasma sheath, which has thickness on the order of the Debye length, necessitates a kinetic description through the Boltzmann equation,

\begin{equation}\label{eq:Boltzmann}
\frac{\partial f_s}{\partial t} = - v \frac{\partial f_s}{\partial x} - \frac{q_s E}{m_s} \frac{\partial f_s}{\partial v} + \left.\frac{\partial f_s}{\partial t}\right\vert_{col} +\left.\frac{\partial f_s}{\partial t}\right\vert_{src}.
\end{equation}

 The particle distribution function $f$ for a given species $s$ is a function of position $x$, velocity $v$, time $t$ and depends on the species mass $m_s$, charge $q_s$, and electric field $E$. Here the equation is solved for 1 spatial and 1 velocity dimension (1X1V) using a discontinuous-Galerkin  (DG) scheme. \cite{juno_discontinuous_2018} The low dimensionality, particularly in velocity space, means that magnetic field effects cannot be accurately captured. As such, to close the system Eq. \ref{eq:Boltzmann} is coupled to Poisson's equation, which gives the electric potential $\phi$ and electric field $E = -\nabla \phi$,

\begin{equation}
    \nabla^2 \phi = \frac{e}{\epsilon_0} (n_e - n_i).
\end{equation}

The particle number densities are given by the zeroth moment of the distribution function $n_s = \int f_s \ dv$. The third term on the right hand side (RHS) of Eq. \ref{eq:Boltzmann} is the collision operator. For this investigation, inter- and intra-species collisions are modeled by a Lenard-Bernstein operator (LBO). \cite{dougherty_model_1964} Aside from electron emission, the walls are assumed perfectly absorbing. As particles are lost to the walls, they carry charge out of the system. The last term on the RHS of Eq. \ref{eq:Boltzmann} is the particle source term which ensures charge conservation in the domain of interest. 

Physically, the emitted electrons originate from the surface of the wall. Therefore, the emitted electron distribution function must be known at the domain boundary. From the definition of the PIEE yield $Y_s = \Gamma_{emit}/\Gamma^{\rightarrow wall}_{s,b}$ and the particle flux for a mono-energetic impacting beam with velocity $v_s'$, $\Gamma^{\rightarrow wall}_{s,b} = v_s' f_s(x_{wall},v_s') \ dv_s$, the emitted electron distribution function for given impact energy can be derived,

\begin{equation}
    \int_{\mathcal{V}_{wall \rightarrow}}{ \frac{m_e}{e} v_e \frac{d Y_s}{d E_e} \ dv_e} = \frac{\int_{\mathcal{V}_{wall \rightarrow}}{v_e f_{emit} \ dv_e}}{\Gamma^{\rightarrow wall}_{s,b}},
\end{equation}

\begin{equation}
    \int_{\mathcal{V}_{wall \rightarrow}}{v_e f_{emit} \ dv_e} = \int_{\mathcal{V}_{wall \rightarrow}}{ v_e \frac{m_e}{e} \Gamma_{s,b}^{\rightarrow wall} \frac{d Y_s}{d E_e} \ dv_e},
\end{equation}

\begin{equation}\label{eq:emitted_dist}
    f_{emit} = \frac{m_e}{e} \frac{d Y_s}{dE_e} \Gamma_{s,b}^{\rightarrow wall}.
\end{equation}

Here $Y_s = \gamma$ for ion impacts and $\delta$ for electron impacts. Note that since the incoming distribution is assumed mono-energetic (delta function), the integration over $v_s$ is dropped. Practically, there are some challenges with this formulation of $f_{emit}$. The first challenge is that $dY_s/dE_e$ is computationally expensive to calculate. It is advantageous to instead use some function $G$, from theory or fitting to experimental data, that can be calculated with reduced computational cost. The fitting function can be defined in the following way,

\begin{equation}
\frac{d Y_s}{d E_e} = \left( \frac{d Y_s}{d E_e} \right)_{max} \frac{dY_s / dE_e}{(dY_s / dE_e)_{max}}, \ \frac{dY_s / dE_e}{(dY_s / dE_e)_{max}} \approx G(E_e).
\end{equation}

Then, the PIEE yield can be rewritten in terms of $G$,

\begin{equation}
Y_s \approx \int \left( \frac{d Y_s}{d E_e} \right)_{max} G(E_e) \ dE_e \Rightarrow \frac{d Y_s}{d E_e} \approx \frac{Y_s}{\int G(E_e) \ dE_e} G(E_e),
\end{equation}
and the emitted electron distribution function at a given impact energy is then approximately given by,

\begin{equation}
f_{emit} \approx \frac{m_e \Gamma_{s,b}^{\rightarrow wall} Y_s}{e \int G(E_e) \ dE_e} G(E_e).
\end{equation}

The other challenge with Eq. \ref{eq:emitted_dist}, is that it is a function of the impacting velocity explicitly and implicitly through $Y_s$ due to the assumption of a mono-energetic impacting distribution function. However, in the continuum-kinetic approach there is only access to the distribution of particle velocities, and this distribution is certainly not mono-energetic for the cases of interest in this work. If the discretized incoming distribution function is treated as a set of mono-energetic beams, then a summation over incoming velocity of Eq. \ref{eq:emitted_dist} gives the total emitted distribution function per impacting species,

\begin{equation}
        f_{emit, \ tot} \approx C G,
\end{equation}
\begin{equation}
        C = \frac{m_e \Gamma_{s}^{\rightarrow wall} \overline{Y}_s}{ e\int G(E_e) \ dE_e},
\end{equation}
\begin{equation}
        \overline{Y}_s = \frac{\sum_{j=1}^{N_{v,s}} v_{s}^j f_{s}^j Y_s(v_{s}^j)}{\Gamma_{s}^{\rightarrow wall}}.
\end{equation}

The average yield $\overline{Y}_s$ uses the particle flux into the wall in the $j^{th}$ velocity space cell as weights. The yield and flux in the $j^{th}$ cell are evaluated using the cell-centered velocity. The total flux $\Gamma_{s}^{\rightarrow wall}$ is given by the first moment of the incoming distribution function over the half of velocity space corresponding to the direction into the wall.

An improved fitting function results in a more accurate representation of $f_{emit}$,

\begin{equation}
 G \rightarrow \frac{dY_s / dE_e}{(dY_s / dE_e)_{max}}, \ f_{emit} \rightarrow \frac{m_e \Gamma_s^{\rightarrow wall} Y_s}{\int G \ dE_e} G. % C G_{fit}. 
\end{equation}

The Chung-Everhart function provides a good fit for the SEE spectrum of various materials. For the IIEE spectrum, the following asymmetric, logarithmic Lorentzian function is found to provide good agreement with both theory and experiment,

\begin{equation}\label{eq:Lorentzian}
\begin{split}
& L(E_s) = \left[ 1 + \frac{\ln^2(E_s/E_0)}{2 \tau^2} \right]^{-1}, \\ \\
& L_a(E_s) = \left\{
\begin{array}{ll}
    L^{\alpha}
    & E_s \leq E_0 \\
    L^{\beta}
    & E_s > E_0 \\
\end{array} 
\right. .
\end{split}
\end{equation}

\begin{figure}[h!]
    \centering
    \includegraphics[width=0.85\linewidth]{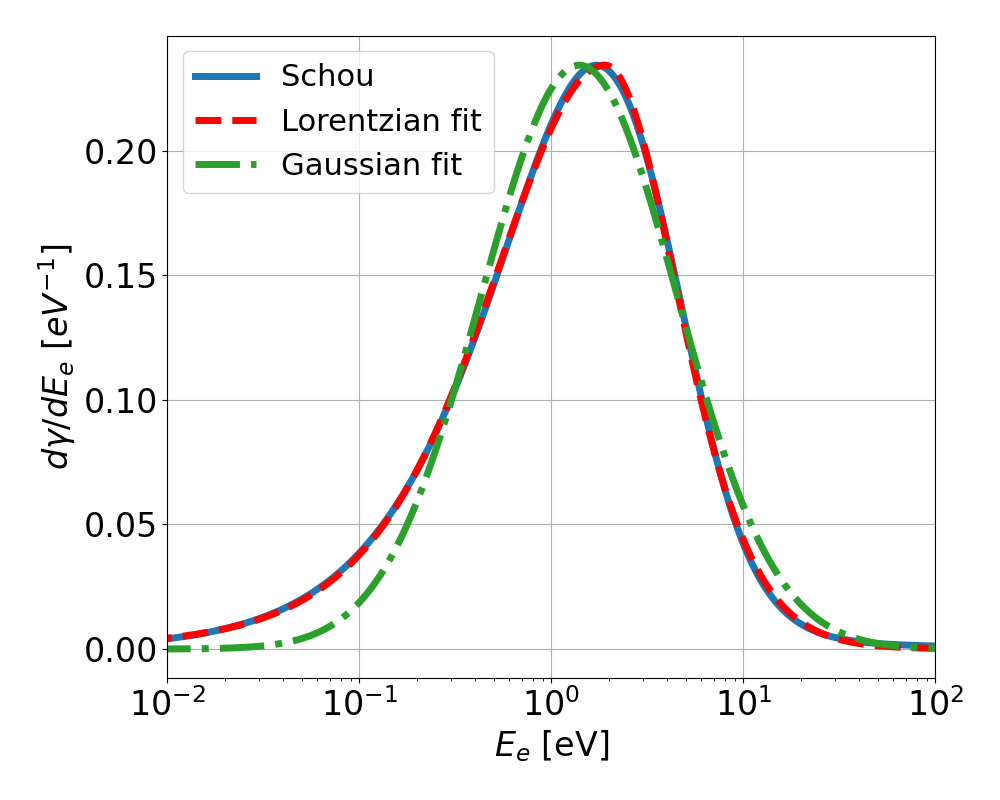}
    \caption{IIEE energy spectrum calculated from Eq. \ref{eq:dgammadE} (blue), the Lorentzian fit (red) and the Gaussian fit \cite{skolar_general_2025} (green).}
    \label{fig:spectrum_fits}
\end{figure}

The terms $\alpha, \ \beta, \ \tau,$ and $E_0$ in Eq. \ref{eq:Lorentzian} are material-dependent fitting parameters. In this work these parameters are found using a least squares fit of Eqs. \ref{eq:dgammadE} and \ref{eq:dgammadE_bind}. The IIEE spectrum calculated for graphite using Eq. \ref{eq:dgammadE_bind} is plotted against the logarithmic Lorentzian and logarithmic Gaussian \cite{skolar_general_2025} fits in Fig. \ref{fig:spectrum_fits}. 

Finally, the full $f_{emit}$ is given by the sum of the contribution from IIEE and SEE. Details of how $f_{emit}$ is applied at the boundary as well as how the reflected populations are handled are given in Ref. \onlinecite{bradshaw_energy-dependent_2024}.

\subsection{Validation of yield and spectrum models}

The PIEE models are tested against experimental data for the materials of interest to determine their validity in this study. To calculate the IIEE yields, the RHS of Eq. \ref{eq:gamma} excluding the ion stopping power, referred to as the ``wall term'', is evaluated. Direct integration of this function can lead to large errors in some materials due to slow convergence. To work around this the Lorentzian-like fitting function, which also approximately gives the shape of the integrand, is integrated using Gauss-Legendre quadrature. The result is then multiplied by the maximum value of the wall term. Finally, the wall term integrated over $E_e$, typically denoted $\Lambda$, is scaled by the ion stopping power to arrive at the energy-dependent yield, $\gamma(E_i) = \Lambda S_{i,e}(E_i)$. Similar to $d\gamma/dE_e$, it is advantageous to fit a function to the ion stopping power data calculated either by SRIM or the LBS, as neither have analytical forms which are easily implemented in the code. Conveniently, the asymmetric, logarithmic Lorentzian is found to provide a good fit for the ion stopping power as well.

\begin{figure*}[t!]
    \includegraphics[width=\linewidth]{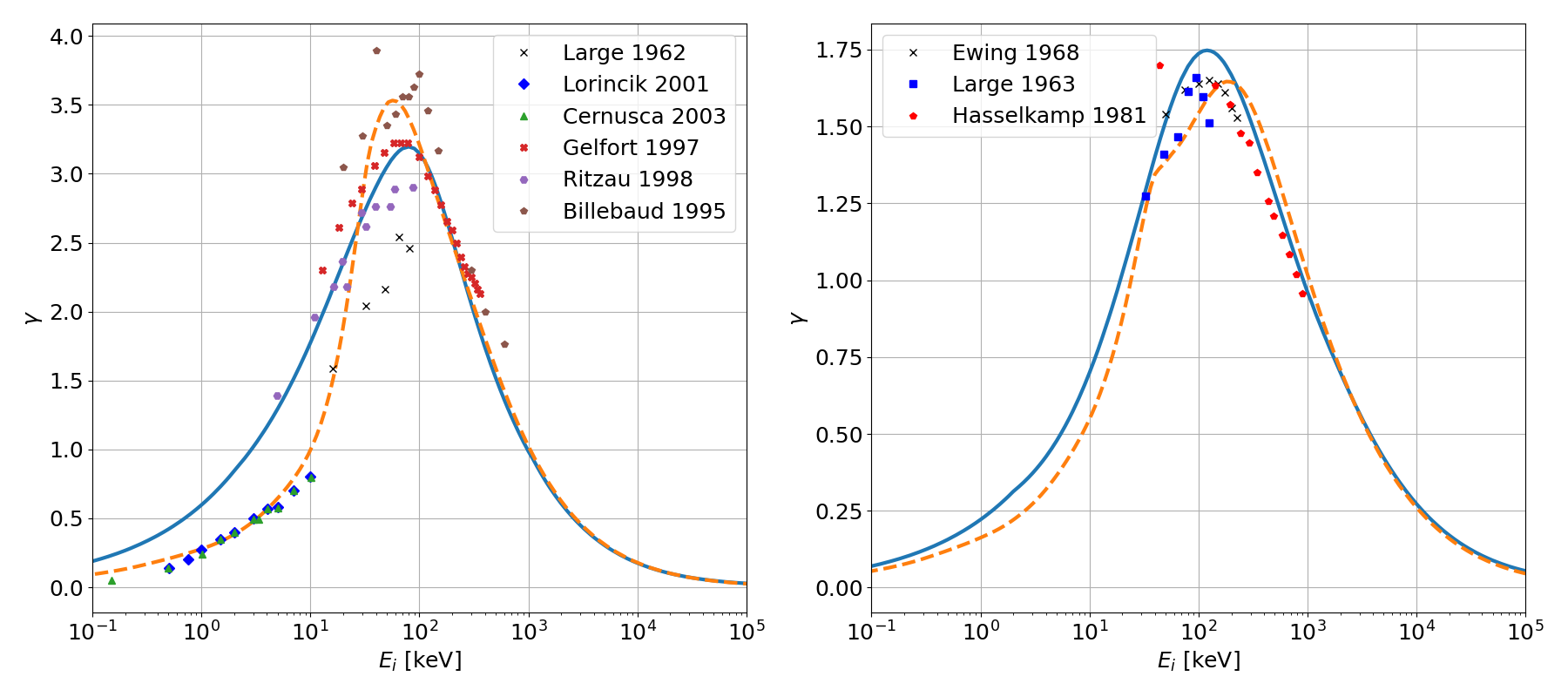}
    \caption{IIEE yield as a function of impacting proton energy. The blue curves are calculated using SRIM and the orange curves with the Lindhard-Bethe sum for the ion stopping power. The experimental data for graphite (left) are from Large and Whitlock, \cite{large_secondary_1962} Lorincik et al., \cite{lorincik_ion_2002} Cernusca et al., \cite{cernusca_ion-induced_2005} Gelfort et al., \cite{gelfort_secondary_1997} Ritzau and Baragiola, \cite{ritzau_electron_1998} and Billebaud. \cite{billebaud_study_1995} Tungsten (right) datasets are from Ewing, \cite{ewing_electron_1968} Large, \cite{large_secondary_1963} and Hasselkamp et al. \cite{hasselkamp_ion_1981}}
    \label{fig:IIEE_yield}
\end{figure*}

The IIEE yield for graphite under proton bombardment calculated in this way is shown in the top left plot in Fig. \ref{fig:IIEE_yield}, which is compared with experimental data collected from several sources. The samples used in many of the experiments are thin carbon foils, which can be graphitic or amorphous allotropes. However the data from Large and Whitlock, \cite{large_secondary_1962} Lorincik et al., \cite{lorincik_ion_2002} and Cernusca et al. \cite{cernusca_ion-induced_2005} are confirmed to be from bulk graphite samples. Even if it is the case that the thin carbon foil samples are not graphitic, SRIM can be used to show that difference in ion stopping power for amorphous carbon and graphite is minimal, and it is therefore reasonable to assume that the same can be said for the yield. Experimental results from Sakamoto et al. corroborate this as well. \cite{sakamoto_stopping_1996} The large variance in the data can be possibly attributed to differences in the density, impurity concentration, and thickness of the sample used. However, the exact conditions of the samples are not known.

It can be seen that $\gamma$ calculated with both SRIM and the Lindhard-Bethe sum fit the data well around the peak. However, the stopping power calculated using SRIM and the LBS differ at low energies. It can be seen in Fig. \ref{fig:IIEE_yield} that the LBS seems to perform better for the data in the range $1 < E_i < 10 $ \qty{}{\kilo\electronvolt}. Based on previous modeling efforts, \cite{skolar_general_2025} it is expected that the peak of the ion distribution that impacts the walls in steady state is within this range, and therefore the LBS is chosen for the presented graphite simulations. For the cases with tungsten, no substantial difference in the yield calculated using either method is observed, and so the curve fitted to SRIM data is used since it is better described by the asymmetric Lorentzian function. Though low energy data near the impact energy range of interest $\sim 1-10$ \qty{}{\kilo\electronvolt} do not exist, good agreement is expected as theory motivates that the yield should still be approximately proportional to the electronic stopping power in this regime. \cite{holmen_direct_1979}

\begin{figure}[h!]
    \centering
    \includegraphics[width=1.0\linewidth]{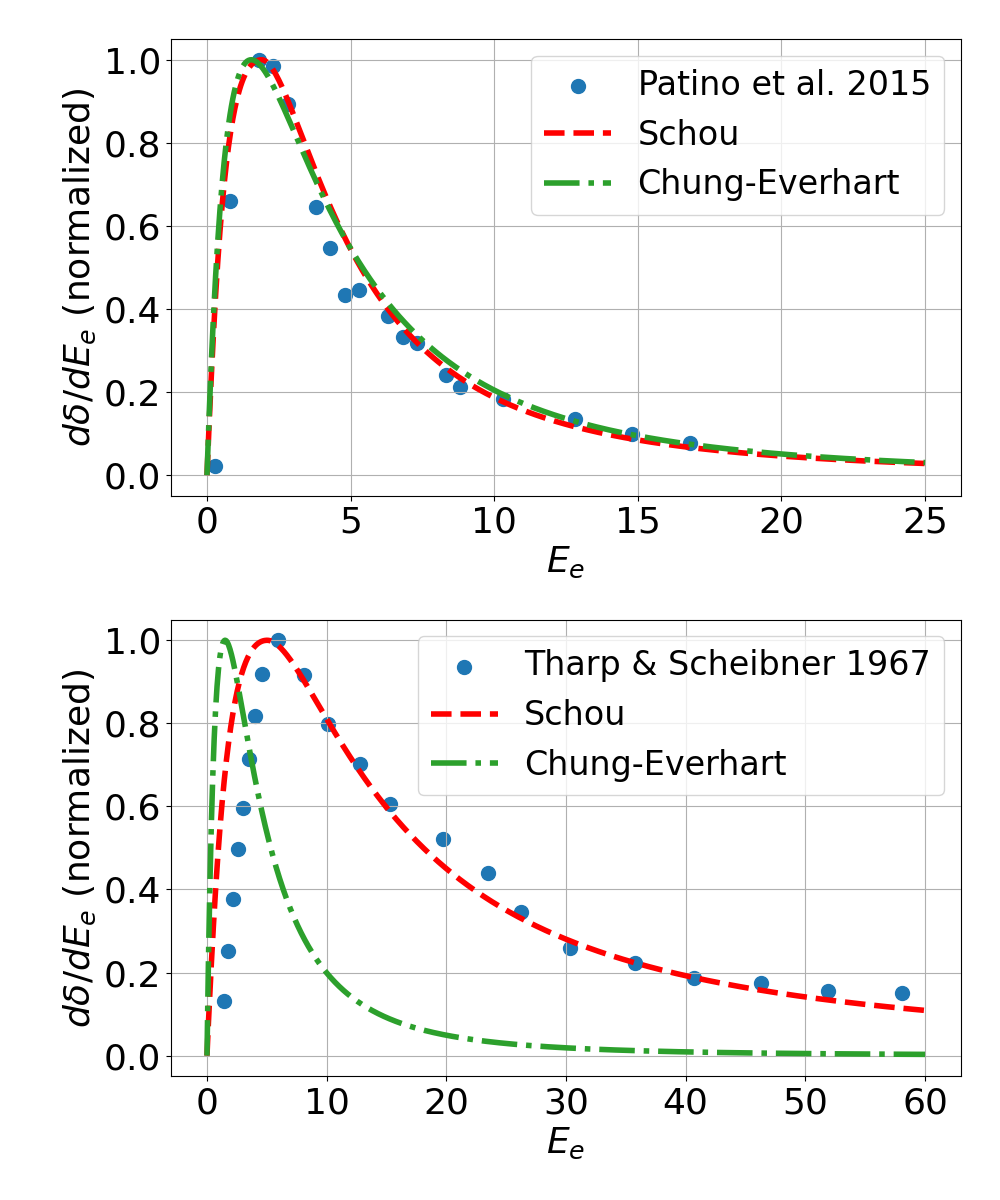}
    \caption{Emitted electron energy spectra for grapihte (top) and tungsten (bottom) using the Schou (red) and Chung-Everhart (green) models. Experimental data are taken from Ref. \onlinecite{patino_analysis_2015} for graphite and Ref. \onlinecite{scheibner_inelastic_1967} for tungsten.}
    \label{fig:graphite_spectrum}
\end{figure}

For SEE, the Furman-Pivi semi-empirical fitting functions are used to obtain the yield curves for graphite and tungsten. Schou's approach in deriving the energy spectra of emitted electrons from transport theory does not explicitly distinguish between the impacting particle charge, \cite{schou_secondary_1987} and thus the shape of the spectrum for ion impacts is expected to be similar to that of electron impacts. The energy spectra calculated with the Schou model are compared against the Chung-Everhart model for tungsten and graphite in Fig. \ref{fig:graphite_spectrum}. The Schou model is shown to have good agreement with experiment for SEE. For graphite both models perform similarly, however the SEE spectrum for tungsten exhibits drastic differences between the two, motivating the use of the Schou model for both IIEE and SEE in the presented simulations.

\begin{table}[h!]
\caption{\label{tab:table1}Comparison of the characteristics from the PIEE spectra calculated using Eq. \ref{eq:dgammadE} with experimental data.  }
\begin{ruledtabular}
\begin{tabular}{lccc}
Material&Emission Type&FWHM [eV]&$E_{peak}$ [eV]\\
\hline
Graphite & theory & 5.06 & 1.71\\
& IIEE\footnotemark[1] & 5.4 $\pm$ 0.4 & 2.0 $\pm$ 0.2\\
& SEE\footnotemark[2] & 5 & 5\\
Tungsten & theory & 17.22 & 4.77\\
% & IIEE & - & -\\
& SEE\footnotemark[2] & 17.44 & 5.69\\
Copper & theory & 10.93 & 3.19\\
% & IIEE\footnotemark[1] & - & -\\
& SEE\footnotemark[3] & 10 & 2.8\\
\end{tabular}
\end{ruledtabular}
\footnotetext[1]{Hasselkamp 1986}
\footnotetext[2]{Tharp and Scheibner 1967}
\footnotetext[3]{Amelio 1970}
\end{table}

Finally, the full-width half maximum (FWHM) and peak energy for the spectra calculated with Eq. \ref{eq:dgammadE} for graphite, tungsten, and copper are compared against data collected from Hasselkamp et al. \cite{hasselkamp_ion-induced_1986} for IIEE, and Amelio \cite{amelio_theory_1970} and Tharp and Scheibner \cite{scheibner_inelastic_1967} for SEE. Though data for the proton-induced spectrum for tungsten do not exist the agreement with the electron-induced spectrum is encouraging for this model. Overall, the PIEE models accurately reflect realistic conditions for an emitting wall and facilitate the presented continuum-kinetic simulations of the emissive sheath.

\section{Setup for simulations}\label{sec:setup}

Given the interest in high pressure plasma discharges with PIEE, the sheaths that form in a Z-pinch fusion device is a natural choice to study given its simple geometry and conditions necessary to drive both significant IIEE and SEE. The simulations presented here take initial conditions from the Fusion Z-pinch Experiment (FuZE) \cite{zhang_sustained_2019} with the ions and electrons initialized as uniform Maxwellians with $T_e = T_i = $ \qty{2}{\kilo\electronvolt} and $n_e = n_i = 1.1 \times 10^{23} $ \qty{}{\m^{-3}}. The domain size is $\pm 256  \lambda_D$ in configuration space and $\pm 4 v_{th,e}$ for electron velocity space and $\pm 6 v_{th,i}$ for ion velocity space. The grid size is $1024_x \times 512_v$ cells for electrons and $1024_x \times 64_v$ for ions. The large configuration space resolution is required to capture sharp gradients close to the walls, and the large electron velocity space resolution is needed to properly resolve the background population and emitted beams.

A source region exists from $\pm 100 \lambda_D$ which ensures conservation of charge and mass. The ion and electron distribution functions in the source region are calculated based on the ion particle flux into the wall, and artificially large collisionality is enforced in this region to maintain a well thermalized population. The mean free path is chosen to be $\lambda_{mfp} = 50 \lambda_D$, i.e. $Kn = 50$. This is chosen because the source region is essentially a scaling down of the bulk plasma from \qty{50}{\cm} to \qty{200}{\micro\meter}. The mean free path of the full Z-pinch is $\sim$ \qty{10}{\cm}, \cite{datta_whole_2024} which when scaled down appropriately becomes $\sim$ \qty{10}{\micro\meter} $\approx 10 \lambda_D$. The collisionality is shaped such that it is highest in the source region and vanishes in the sheath with profile given by,

 \begin{equation}\label{eq:collision_profile}
 \begin{split}
 & H(x) = h(x) + h(-x)  - 1, \\ \\
 h(x) & = \left[ 1 + \exp{\left( \frac{x}{12 \lambda_D} - \frac{16}{3} \right)} \right]^{-1}.
 \end{split}
 \end{equation}

 The profile scales the collision frequency, which is approximated as $\nu_{ss}=v_{th,s}/\lambda_{mfp}$ with $v_{th,s}$ being the species thermal velocity. For cross-species collisions, the collision frequencies are given as $\nu_{ei} = \nu_{ee}$ and $\nu_{ie} = \nu_{ee} \sqrt{m_e/m_i}$. 

The applied voltage to the electrodes is varied with Dirichlet boundary conditions for Poisson's equation. The cathode is held fixed at \qty{0}{\kilo\volt} while the anode potential is varied from \qty{0}{\kilo\volt} to \qty{10}{\kilo\volt} in increments of \qty{2}{\kilo\volt}. A sheath naturally develops due to higher electron mobility and the initial charge imbalance launches Langmuir waves which gradually damp out. \cite{cagas_continuum_2017} The spectrum and yield curves calculated for graphite and tungsten are used for calculating the emitted distribution function at each timestep. The simulations reach a quasi-steady state after 10,000 plasma oscillation periods.

The net plasma current can be calculated using,

 \begin{equation}
 I = \pi r^2e \langle \Gamma_i(x) - \Gamma_e(x) \rangle
 \end{equation}
 where $r$ is the pinch radius, taken to be \qty{3}{\milli\meter} as observed in experiments. \cite{zhang_sustained_2019} The angled brackets represent the domain-averaged value. This formula assumes that the pinch takes the shape of a cylinder with constant cross-section and that the current density remains constant in the radial direction. The ion and electron temperatures are related to the second moment of their respective distribution functions by,

 \begin{equation}\label{eq:moment_temp}
 \begin{split}
 & M_{2,s} = \int_{-\infty}^{\infty} v^2 f_s \ dv, \\ \\
 & T_s = \frac{m_s}{n_s} \left( M_{2,s} - \frac{\Gamma_s^2}{n_s} \right).
 \end{split}
 \end{equation}

Similarly, the heat flux is related to the third moment by,

 \begin{equation}\label{eq:moment_heat_flux}
 \begin{split}
 & M_{3,s} = \int_{-\infty}^{\infty} v^3 f_s \ dv, \\ \\
 q_s = m_s & \left[ \frac{1}{2} M_{3,s} - \frac{\Gamma_s}{n_s} \left( \frac{3}{2} M_{2,s} -\frac{\Gamma_s^2}{n_s} \right) \right].
 \end{split}
 \end{equation}

Three cases are run while sweeping through the applied voltage to the electrodes: no emission, IIEE only with tungsten and graphite electrodes, and IIEE and SEE with graphite electrodes.

\section{Results and discussion}\label{sec:results}

The transport properties of the unmagnetized, non-emissive plasma sheath are largely governed by the potential gradient, which acts as a high energy filter for electrons and generates a drifting ion population. \cite{stangeby_plasma_2000} The fluid theory discussed in Sec. \ref{sec:SheathTheory} shows a strong dependence of the particle and heat fluxes on the sheath potential drop for this reason. The inclusion of emitted electrons in the sheath is thought to decrease the potential gradient, via accumulation of negative space charge, leading to more severe heat and particle loads to the walls. 

\begin{figure}[h!]
    \centering
    \includegraphics[width=1.0\linewidth]{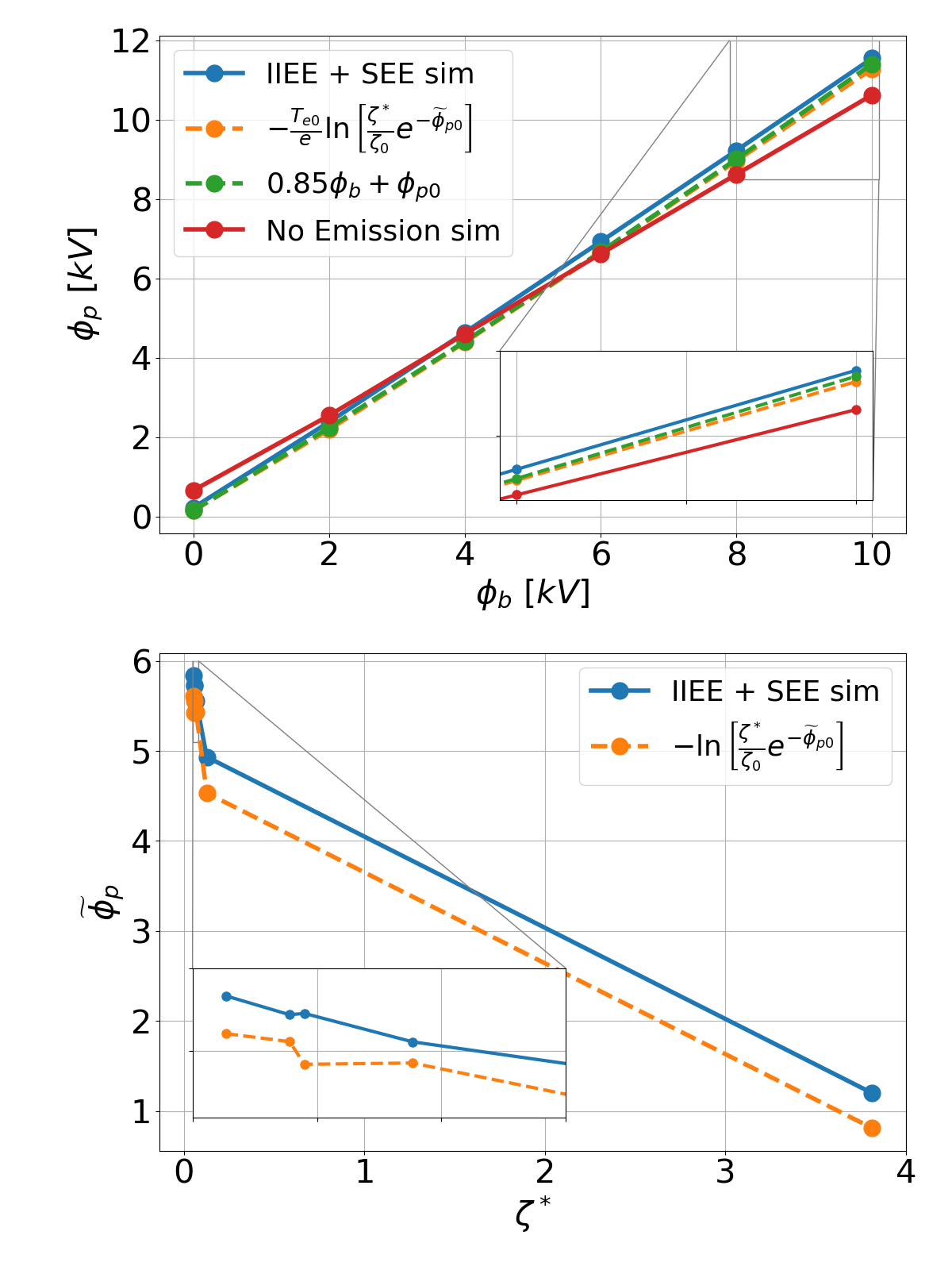}
    \caption{Plasma potential as a function of bias potential (top) and general effective yield (bottom).}
    \label{fig:phi_p}
\end{figure}

Fig. \ref{fig:phi_p} presents the plasma potential from the kinetic simulations as a function of the applied bias and of the general effective yield. The plasma potential from the simulations is defined as the average of the potential at the source edges, $x = \pm 100 \lambda_D$, on the anode and cathode sides. The potential derived from fluid theory, Eq. \ref{eq:plasma_potential_Emit}, shows excellent agreement with the simulation results, which both follow a linear trend with the applied voltage. 

\begin{figure}[h!]
    \centering
    \includegraphics[width=1.0\linewidth]{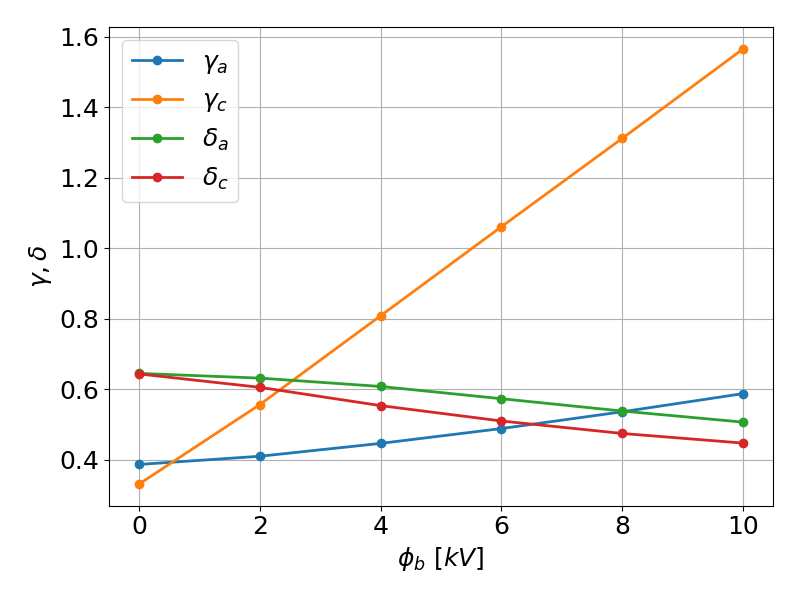}
    \caption{Ion- ($\gamma$) and electron- ($\delta$) induced yields. Note that the electron-induced yields at the anode and cathode include both true secondary and reflected electrons.}
    \label{fig:yields}
\end{figure}

Though the IIEE yield increases with the bias potential, as is indicated in Fig. \ref{fig:yields}, the total SEE yield decreases and the effective yield at the anode decreases as well. Still, being that there is some emission, a question naturally arises: why does the plasma potential increase faster than cases without emission? The reason comes from the effect shown in Ref. [\onlinecite{levko_electron_2014},\onlinecite{skolar_general_2025}] in which emitted electrons are accelerated through the sheath electric field into the presheath where they preferentially give up their energy to bulk electrons through collisions. The emitted electron heating effect drives the ion-electron temperature ratio down thereby increasing the plasma potential. The effect is primarily driven by emitted electrons from the cathode due to the much larger potential drop in the cathode sheath.

The emitted electron heating dominates over the influence of the emitted electron space charge on the potential because the potential is directly proportional to both $T_{e0}$ and $\ln{(1/\zeta^*)}$, as shown in Eq. \ref{eq:plasma_potential_Emit}. Note that $T_{e0}$ here is taken to be the average of the temperature values at the source edges.

\begin{figure}[h!]
    \centering
    \includegraphics[width=1.0\linewidth]{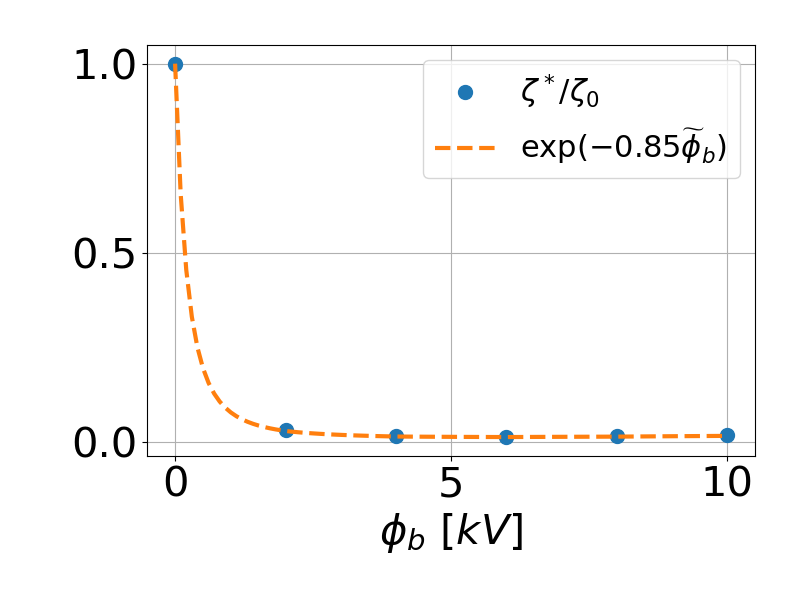}
    \caption{Normalized general effective yield as a function of bias potential (blue) and exponential fit (dashed organge).}
    \label{fig:eff_yield_fit}
\end{figure}

Plotting the $\zeta^* / \zeta_0$ term in Eq. \ref{eq:plasma_potential_Emit} against $\widetilde{\phi}_b$, as shown in Fig. \ref{fig:eff_yield_fit}, one notices a simple exponential can be fit of the form $\exp{(-\kappa \widetilde{\phi}_b)}$. Plugging this fitting function back into Eq. \ref{eq:plasma_potential_Emit} provides a simple expression for the plasma potential,

\begin{equation}
    \widetilde{\phi}_p = \kappa \widetilde{\phi}_b + \widetilde{\phi}_{p0}.
\end{equation}

For graphite the fitting parameter is found to be $\kappa = 0.85$. The advantage of this fitting function is that it does not require knowledge of how the emission characteristics of the material change with the applied bias to obtain the emission-corrected plasma potential.

 In Fig. \ref{fig:phi_p} the reduced plasma potential $\widetilde{\phi}_p = e \phi_p / T_{e0}$ is plotted against $\zeta^*$, generally showing a decreasing trend with increasing $\zeta^*$ as expected. The potential is shown to have a slightly non-monotonic relationship with the general effective yield because, as shown in Eq. \ref{eq:plasma_potential_Emit}, the bias potential and emitted electron heating compete with the emitted electron space charge in their influence on the potential.

 \begin{figure}[h!]\label{fig:phi_profiles}
    \centering
    \includegraphics[width=1.0\linewidth]{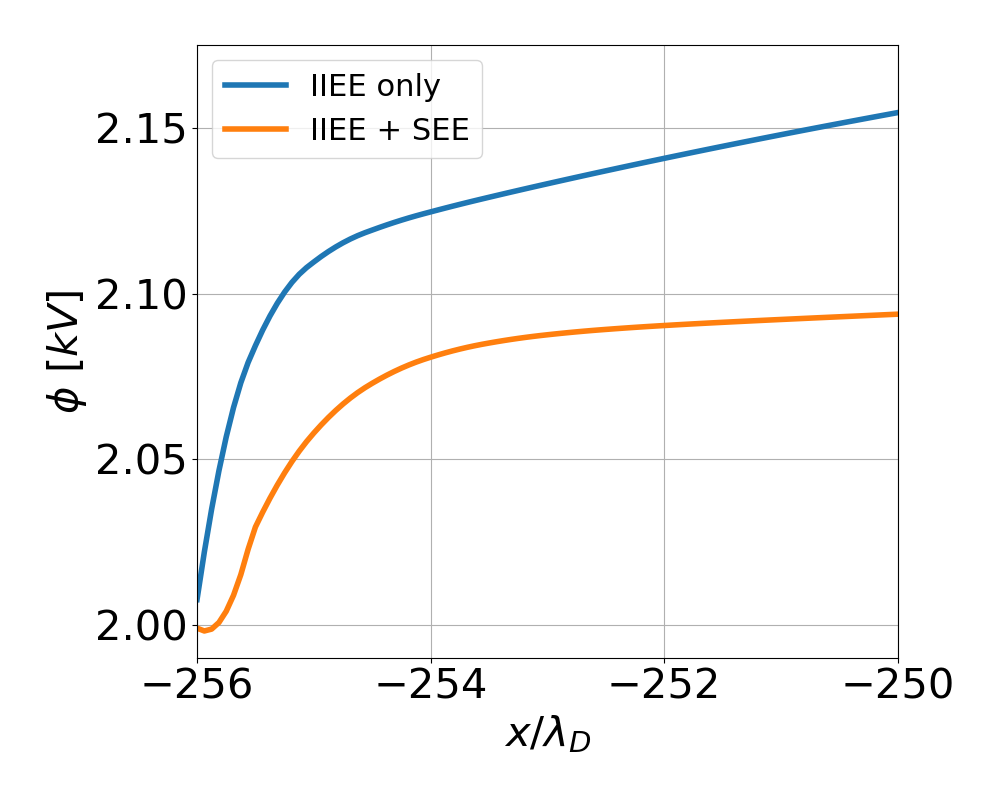}
    \caption{Potential profiles in the sheath at 2kV biased graphite anode.}
    \label{fig:phi_profiles}
\end{figure}

 An interesting prediction from fluid theory is that IIEE has a lower impact on the potential relative to SEE. The implication of this prediction is that the structure of the sheath depends primarily on the SEE yield. The potential profiles from the simulations shown in Fig. \ref{fig:phi_profiles} validate this as at 2 kV applied bias, the anode transitions from a classical to an SCL sheath. The case with only IIEE is compared against the case with both types of PIEE. A virtual cathode forms only in the PIEE case, which reflects low energy emitted electrons back to the wall. The virtual cathode potential well has a depth of $\approx$ \qty{2}{\volt}, which is small compared to the sheath potential drop, however it is important to recall most emitted electrons will have energy on the order of 1 eV as shown in Fig. \ref{fig:graphite_spectrum}.
 
 The sheath remains SCL except for the 10 kV case where the SEE yield dips below the IIEE yield, shown in Fig. \ref{fig:yields}. The depth of the virtual cathode decreases with increasing bias potential, almost halving between the 2 and 8 kV cases, again showing the relative importance of SEE over IIEE. In both cases the cathode sheaths remain classical, due to the bias potential being primarily accounted for in the cathode sheath, as noted in Ref. \onlinecite{campanell_two_2025}. This leads to a larger electric field that makes it difficult to achieve the space-charge limit by PIEE alone. 

\begin{figure}[h!]
    \centering
    \includegraphics[width=1.0\linewidth]{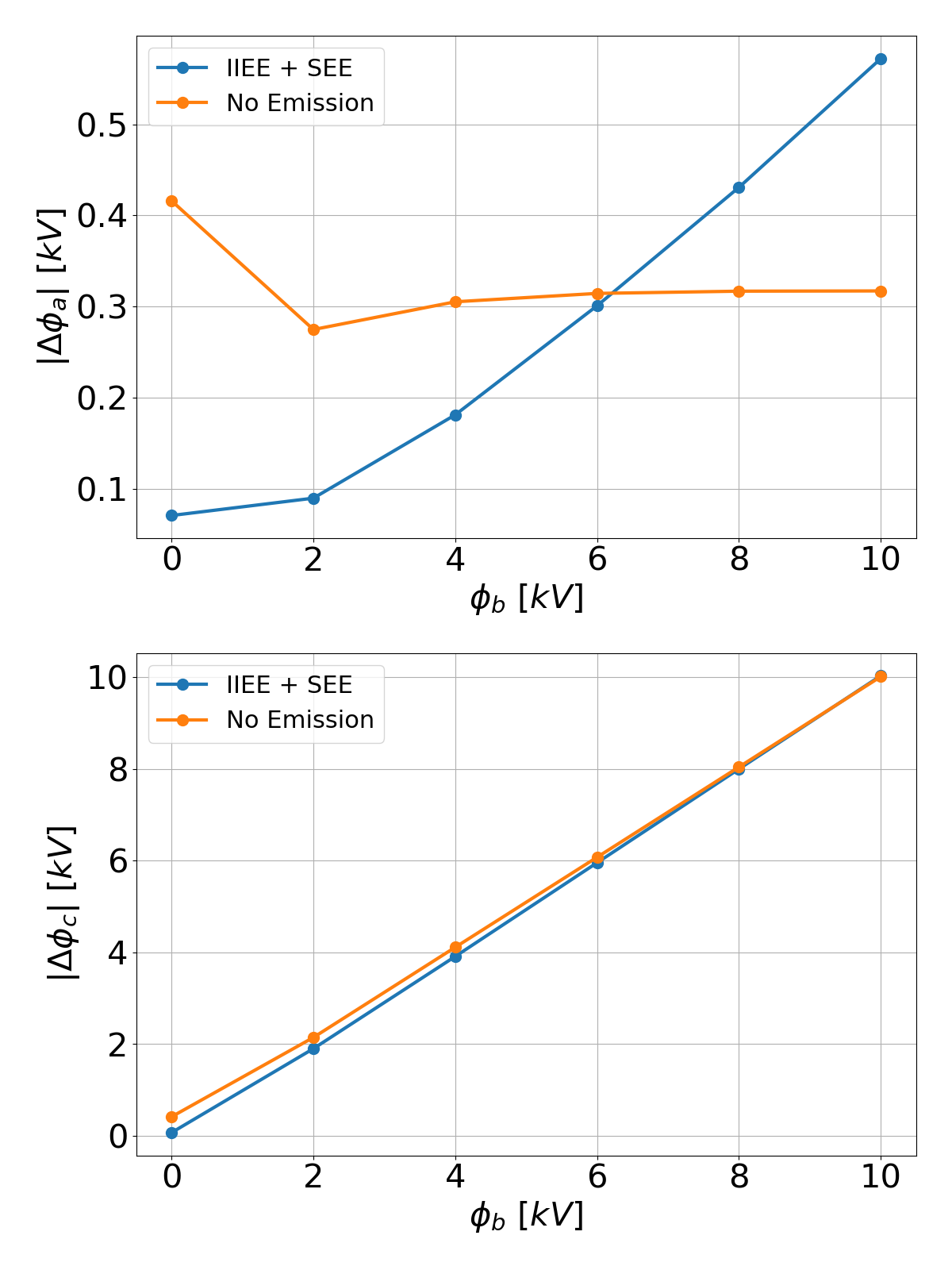}
    \caption{Potential drop across the sheath as a function of bias potential at the anode (top) and cathode (bottom).}
    \label{fig:phi_drop}
\end{figure}

\begin{figure}[h!]
    \centering
    \includegraphics[width=1.0\linewidth]{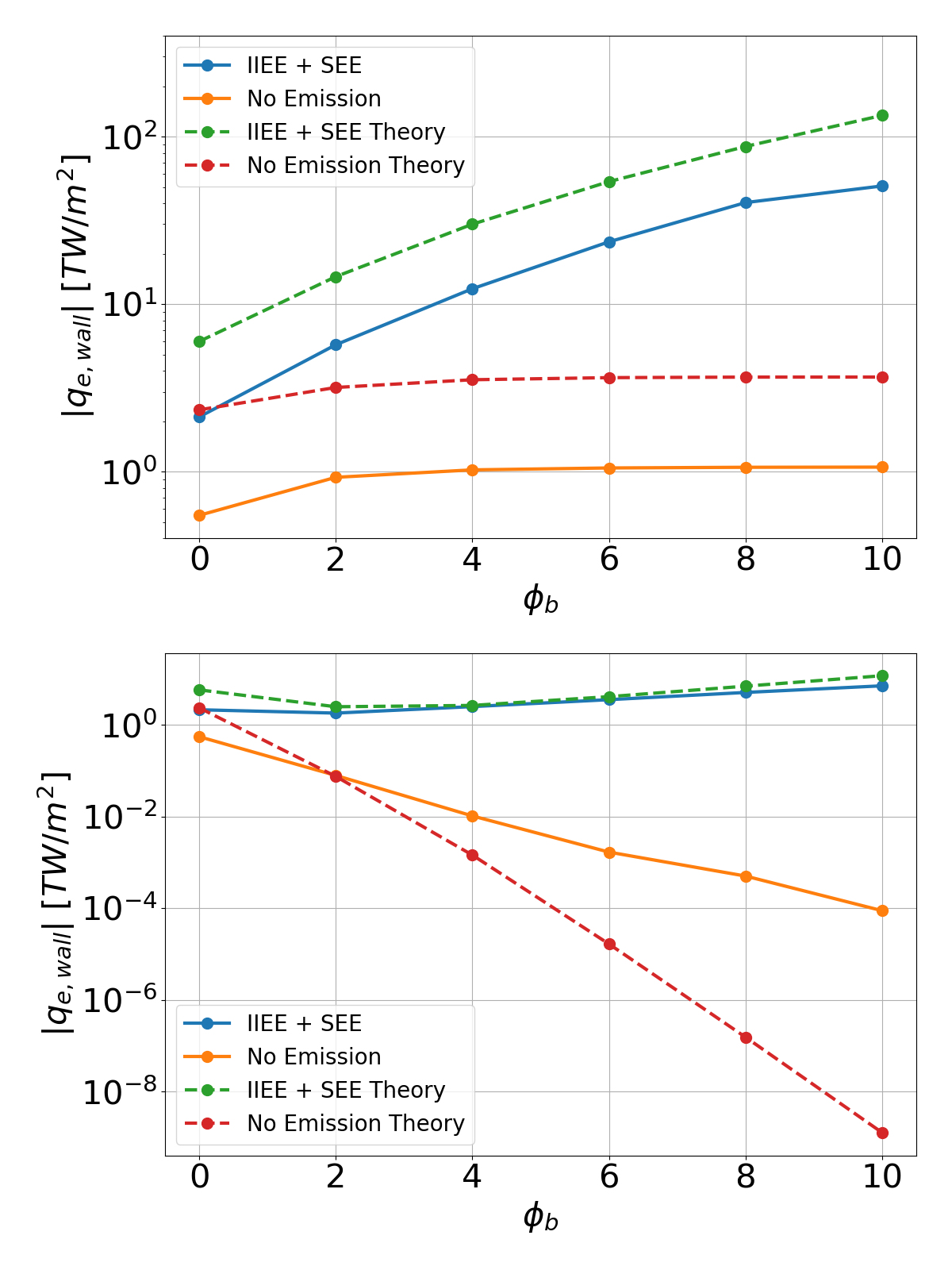}
    \caption{Net electron heat flux at the anode (top) and cathode (bottom) for the cases with and without PIEE. Dashed curves are calculated using fluid theory from Sec. \ref{sec:SheathTheory}.}
    \label{fig:heat_flux_PIEE_vs_NE}
\end{figure}

Fig. \ref{fig:heat_flux_PIEE_vs_NE} shows the electron heat flux at the anode and cathode plotted against the bias potential. The ion heat flux is found to be negligible in all cases and so only discussion of the electron heat flux is presented here. In addition, discussion of the effects of emission on the heat flux will be with regards to the anode sheath to isolate from effects of the applied voltage. As the bias potential increases, the IIEE yield increases while the SEE yield decreases at both electrodes. For the emissive cases, the potential difference across the anode and cathode sheaths increase with increasing applied bias, shown in Fig. \ref{fig:phi_drop}, which suggests there should be a reduction in heat flux to the wall from conventional sheath theory. 

However, the plasma potential is not the only property to be impacted by emitted electrons. As mentioned previously, electron heat flux has been predicted to be proportional to the electron temperature at the sheath edge both from fluid and kinetic theory. \cite{stangeby_plasma_2000,tang_bohm_2017} The increasing potential drop across the sheath drives further acceleration of emitted electrons, leading to larger temperatures at the sheath edge, which is defined here as where quasineutrality is perturbed by 0.1\%, $|(n_i-n_e)/(n_i+n_e)|\geq0.001$.

Viewed from a kinetic standpoint, a higher temperature at the sheath edge corresponds to a broader electron distribution function with enhanced tail populations. As electrons traverse the sheath, absorption of high energy electrons truncates the distribution in velocity space. At the wall only the half of the distribution moving towards the wall remains. The distribution is further truncated since low energy electrons are unable to overcome the potential barrier, and are reflected before they reach the wall. These effects produce an asymmetric distribution about the local drift velocity. Because the heat flux is defined as being proportional to the third central moment of the distribution function, this increased asymmetry in a truncated, broader distribution leads to a larger heat flux to the wall.

\begin{figure}[h!]
    \centering
    % \hspace*{-0.7cm} 
    \includegraphics[width=1.0\linewidth]{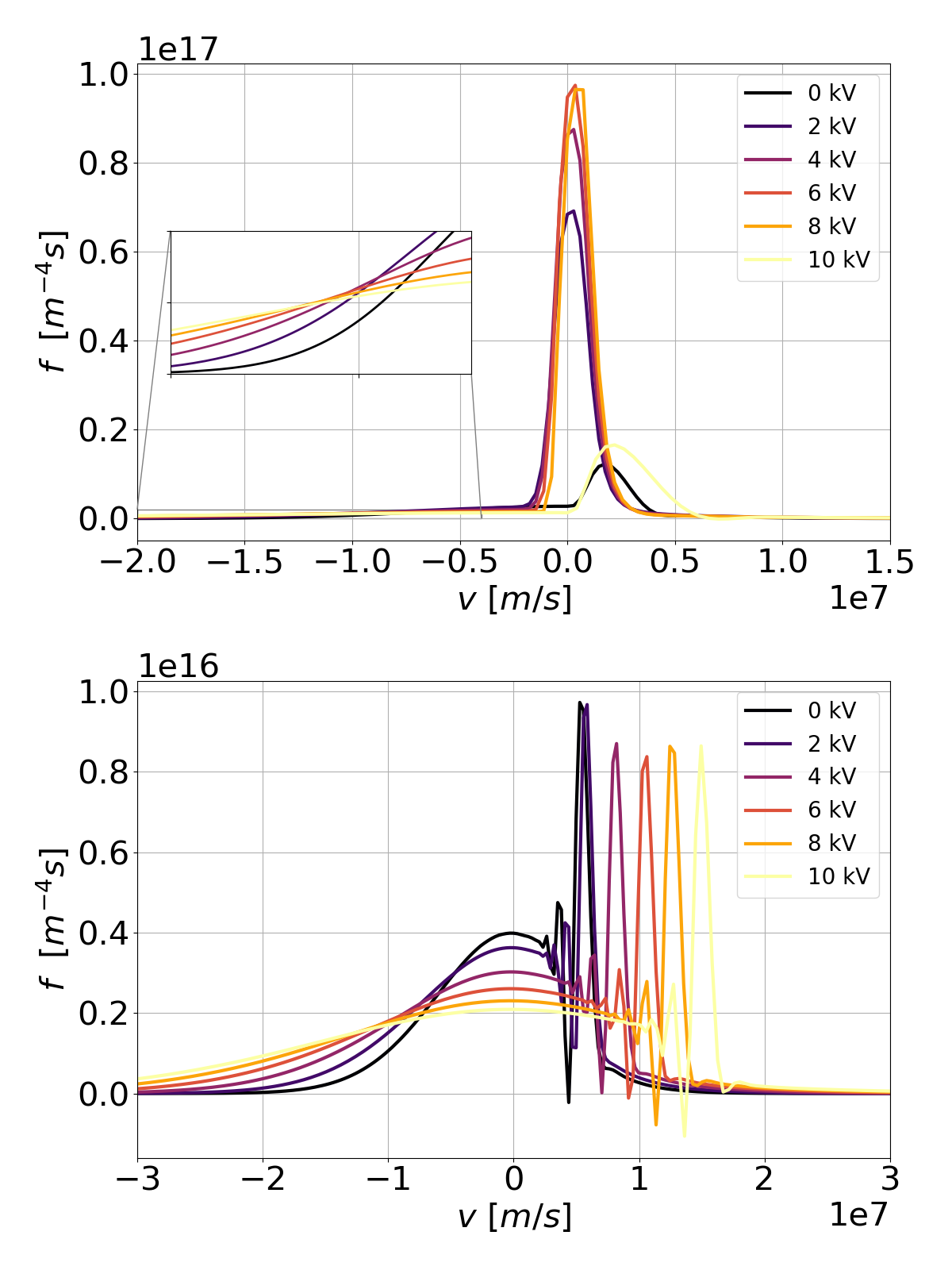}
    \caption{Electron distribution function at the anode (top) and at the anode sheath edge (bottom) at increasing bias potentials for the IIEE and SEE case.}
    \label{fig:dist_fun_anode}
\end{figure}

The electron distribution function at the anode and anode sheath edge can be examined in Fig. \ref{fig:dist_fun_anode} to gain further insight into the origin of increasing heat flux with PIEE. Note that for the distributions at the sheath edge, numerical oscillations are observed due to sharp velocity space gradients. While the effective yield at the anode is found to decrease with bias potential, the flux of emitted electrons increases. The broadening of the distribution function then coincides with an increase in both the emitted flux and sheath potential drop. Thus, more energy from emitted electrons is collisionally transferred into the background electron population in the presheath, creating enhanced tails of the distribution function. The asymmetry in the distribution function at the wall, and therefore the heat flux into the wall, is due to the balance between the enhanced tail population in the background electrons and the emitted beam itself. 

The emitted beam in the \qty{10}{\kilo\volt} case decreases significantly because the steady state sheath becomes classical late in time after initially transitioning to SCL. After the collapse of the SCL sheath, the virtual cathode, which partially confines emitted electrons, is destroyed allowing for a free streaming emitted beam and lower beam density at the wall. The transition from SCL to classical likely occurs due to increasing flux of electrons into the wall and decreasing SEE yield, which has been demonstrated to play a dominant role in the transition between classical and SCL sheaths.

The theory fully describing this transition and the associated conditions constitutes future work. However, preliminary models suggest the effective yield achieved in the \qty{10}{\kilo\volt} case is close to the critical value for a transition between an SCL and classical sheath. Previous modeling efforts \cite{hobbs_heat_1967,takamura2004} have neglected ion thermal effects which are expected to play a substantial role in determining this critical effective yield for fusion conditions.

\begin{figure}[h!]
    \centering
    \includegraphics[width=1.0\linewidth]{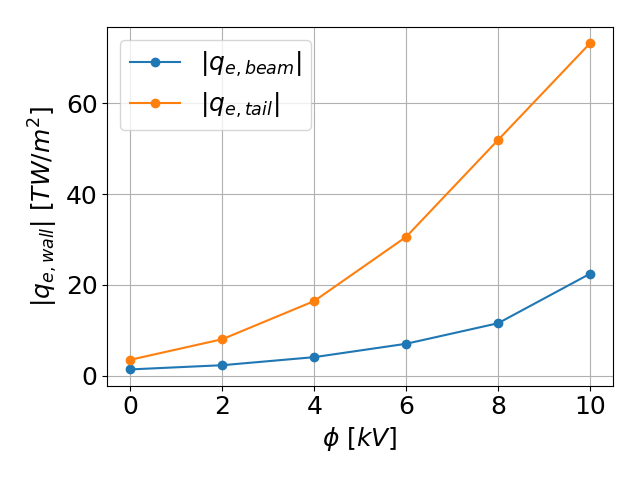}
    \caption{Emitted beam and tail contributions to the electron heat flux at the anode.}
    \label{fig:heat_flux_beam_vs_tail}
\end{figure}

Shown in Fig. \ref{fig:heat_flux_beam_vs_tail}, the beam and tail contributions to the heat flux into the anode are approximately captured by the third central moment in the positive and negative half of velocity space, respectively. The tail component dominates the beam component, as required for net heat flow into the wall because the beam carries heat away from the wall and into the plasma.

\begin{figure}[h!]
    \centering
    \includegraphics[width=1.0\linewidth]{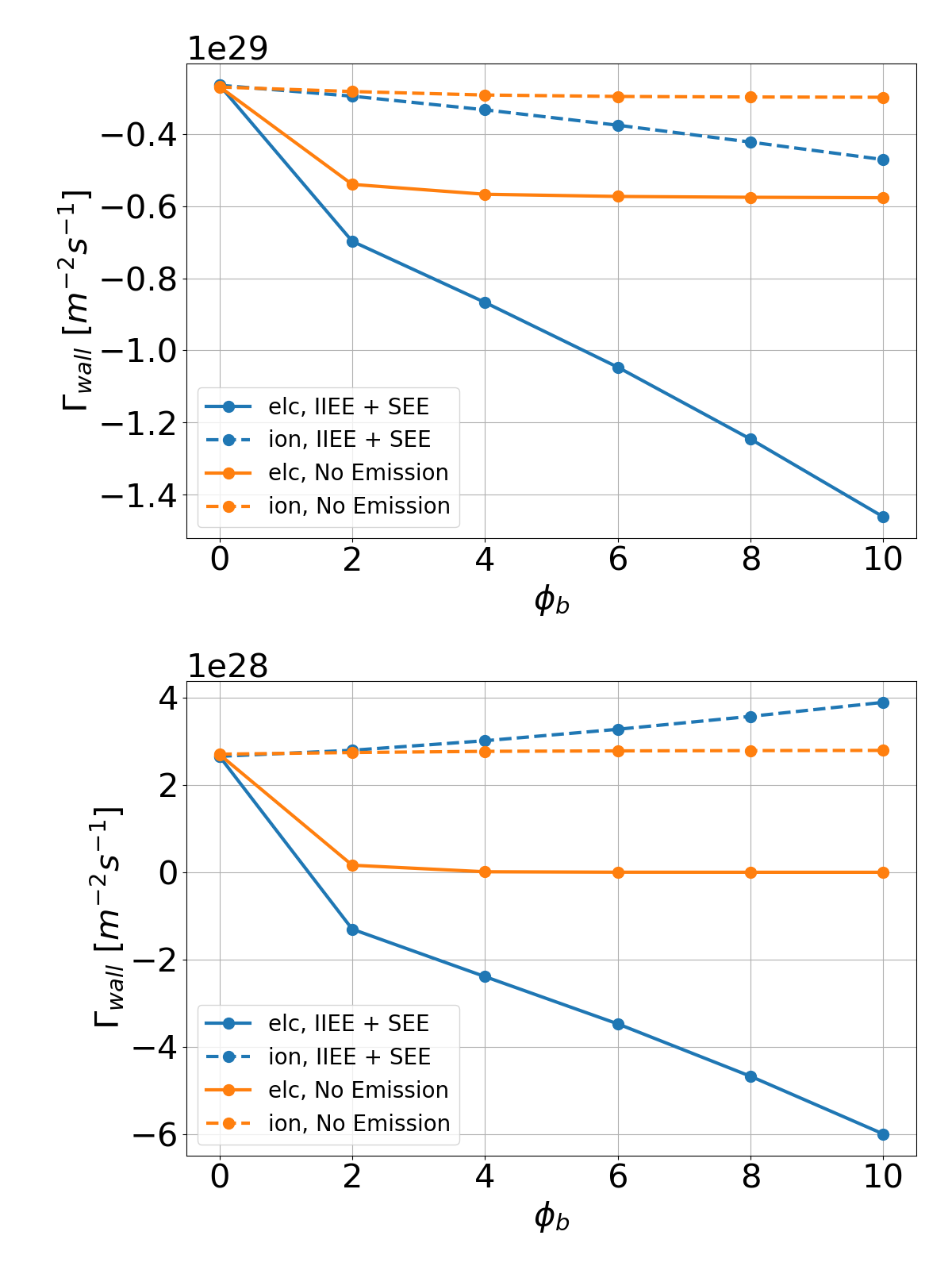}
    \caption{Electron and ion particle flux at the anode (top) and cathode (bottom) for the cases with and without PIEE. A negative flux denotes particles moving towards the anode and positive towards the cathode.}
    \label{fig:particle_flux}
\end{figure}

Like the heat flux, the behavior of the particle flux depends strongly on which side of the discharge one is examining. Fig. \ref{fig:particle_flux} compares the fluxes of ions and electrons at the anode and cathode with and without PIEE. At both electrodes for the cases without emission, the electron particle flux saturates as fluid theory predicts. The ion flux remains constant as the bias potential increases suggesting the product of the sheath edge density and Bohm speed do not vary significantly, which is found to be the case from the simulation data. When PIEE is introduced, the electron flux increases with slope consistent with fluid theory. \cite{skolar_general_2025}

\begin{figure}
    \centering
    \includegraphics[width=1.0\linewidth]{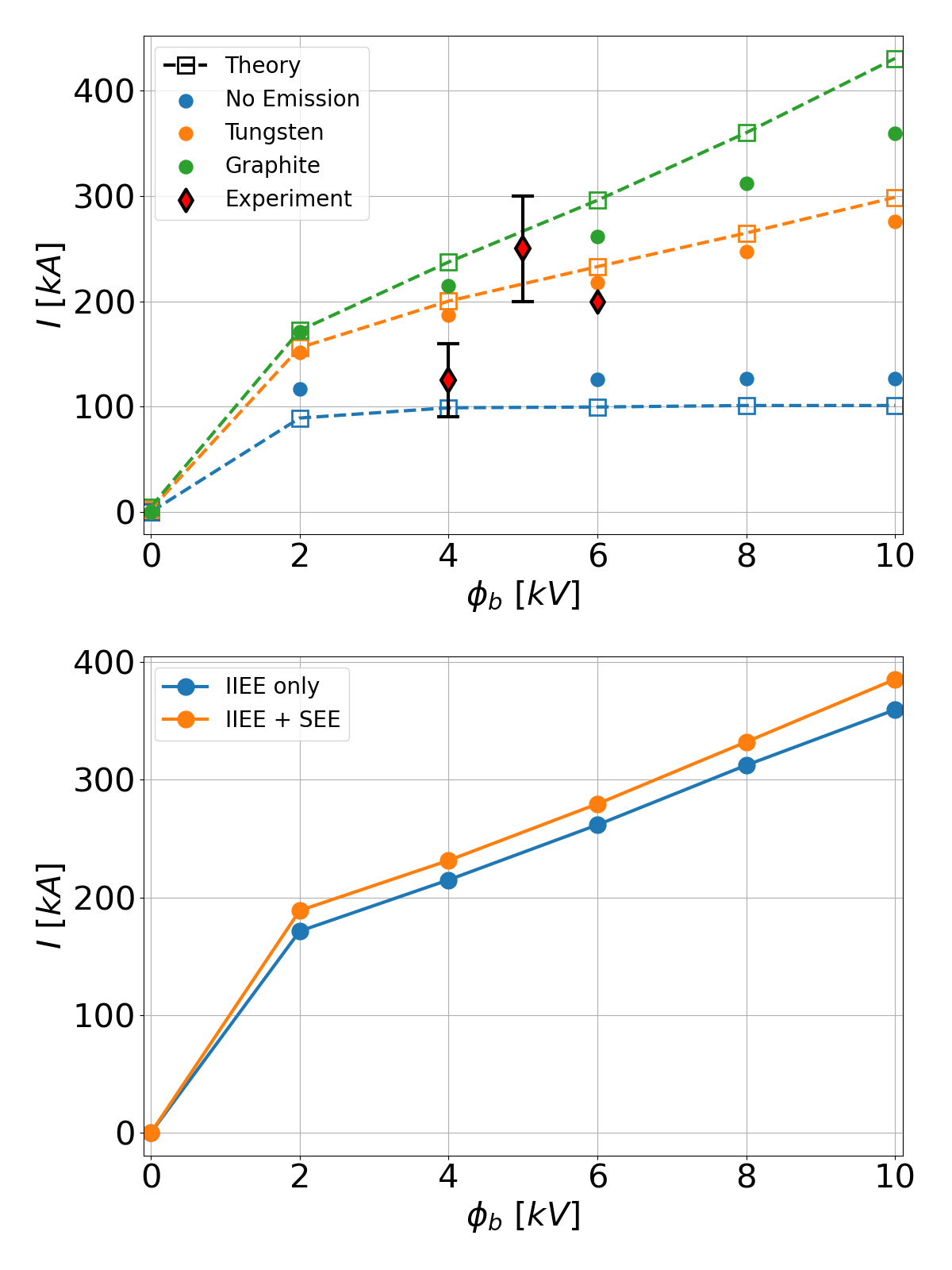}
    \caption{Pinch current for the IIEE only cases (top) and PIEE cases (bottom). Experimental data are taken from Ref. \onlinecite{Claveau2019,zhang_sustained_2019}.}
    \label{fig:pinch_current}
\end{figure}

The pinch current is found from the difference of the ion and electron particle fluxes. The top of Fig. \ref{fig:pinch_current} shows the results of the cases with IIEE only compared against the no emission cases. The trends of the current with bias potential are found to agree with Eq. \ref{eq:pinch_current}. Here, the theoretical pinch current is found by averaging the anode and cathode currents, which tend to be very similar, and assuming a Bohm speed of the form $c_s = \sqrt{(T_e + 3 T_i)/m_i}$. \cite{tang_critical_2016} The temperatures used in calculating the Bohm and average electron speeds are taken to be an average of the sheath edge values, however recall the plasma potential is calculated with the average of the source edge values. In this way, the full potential drop experienced by ions and electrons is taken into account while preserving continuity. The pinch current for the cases with IIEE exhibit the expected linear trend based on the linear relationship between $\gamma$ and $\phi_b$ shown in Fig. \ref{fig:yields}. The current for the cases with graphite electrodes and both IIEE and SEE is also shown in Fig. \ref{fig:pinch_current}, which is slightly increased over the IIEE cases for non-zero bias potentials but has the same slope. 

Limited experimental data is available for the pinch current at varying applied voltages on the FuZE device. However, from the data that is accessible the simulation results are found to provide good agreement, particularly when PIEE is properly accounted for. 

\section{Conclusion}\label{sec:concl}

The presented continuum-kinetic sheath simulations with PIEE demonstrate the importance of emitted electron heating on macroscopic transport properties and the structure of the sheath for fusion plasma. The heat and particle fluxes increase in spite of an increasing potential drop across the sheath. The magnitudes of the heat flux agree with previous modeling efforts \cite{thompson_electrode_2023,datta_whole_2024} and the plasma current calculated from the particle fluxes match fairly well with experimental measurements. \cite{Claveau2019,zhang_sustained_2019} The total heat load on the electrodes is found to be on the order of megawatts, which while large is not uncommon for fusion devices. 

Fluid theory is revised to account for both ion- and electron-induced emission and shows agreement with the simulated plasma properties. This revised theory predicts that SEE has a greater influence on the sheath structure than IIEE, which is validated by the kinetic simulations showing a transition to an SCL regime when SEE is included under the same conditions. Magnetized sheath simulations \cite{Moritz2018-eq,Skolar2026-fh} with PIEE will allow for better estimates of the quantities of interest, and to explore other applications such as tokamaks and stellarators.

For the Z-pinch specifically, a high pinch current is desirable as the fusion output has been shown have an eleventh power relationship with the current. \cite{shumlak_z-pinch_2020} However, increased emission is a doubled-edged sword as emitted electron heating increases the ion-electron temperature ratio, leading to larger radiative losses via bremsstrahlung, and increased heat and particle deposition on the electrodes, resulting in larger erosion rates and flux of impurities into the plasma. 

An algorithm similar to the one used in the IIEE boundary condition can be used to estimate the flux of sputtered and sublimated neutrals from the walls provided accurate, energy-dependent yields. The empirical fits for sputtering and sublimation yields from Ref. \onlinecite{roth_erosion_1991} and wall temperature from Ref. \onlinecite{thompson_electrode_2023} are used to calculate a total flux of neutrals from the cathode of $1.78 \times 10^{28}$ \qty{}{\meter^{-2}\second^{-1}} for graphite and $8.71 \times 10^{28}$ \qty{}{\meter^{-2}\second^{-1}} for tungsten at \qty{10}{\kilo\volt}, which is consistent with the experimental measurement of $3.62 \times 10^{28}$ \qty{}{\meter^{-2}\second^{-1}} on the FuZE device. \cite{beers_cathode_2022} An approximate erosion rate can then be easily obtained and for graphite is found to be $\sim$ \qty{10}{\gram / \s} . Assuming power plant operating conditions from Ref. \onlinecite{thompson_electrode_2023}, namely a pulse rate of \qty{10}{\hertz} and plasma lifetime of \qty{200}{\micro\s}, the total eroded material from the cathode alone is found to be \qty{1.7}{\kilo\gram} per day of continuous operation. For tungsten electrodes this is $\sim 100$ times larger.

Finally, Levko and Krasik \cite{levko_electron_2014} discuss the excitation of a streaming instability from cathode emissions in a plasma discharge. Above a threshold pressure, the instability was no longer excited due to a much larger ratio between bulk electron and emitted beam densities. However, their model lacked emission from the anode which, in the case of sufficiently low collisionality in the bulk, can lead to the excitation of the two-steam instability even at fusion-relevant pressures. \cite{bradshaw_effects_2025}
Though not discussed in detail here, the excitation of such an instability is observed when SEE is taken into account and the Bhatnagar-Gross-Krook (BGK) collision operator is used. The BGK operator relaxes the suppression of the high energy emitted electron beams in the presheath by the LBO, allowing them to interact in the bulk. The instability can further exacerbate heat and particle fluxes to the walls, and a more thorough analysis with a Fokker-Planck operator is planned in future publications.

\begin{acknowledgments}

The authors acknowledge the Hyak supercomputer system at the University of Washington for providing computational resources for the simulations presented in this paper.
\end{acknowledgments}

\section*{Data Availability Statement}

The presented simulations were performed and can be reproduced with the open-source software Gkeyll. Information for obtaining, installing, and running Gkeyll may be found on the documentation site: https://gkeyll.readthedocs.io.

\appendix

% \section{Appendixes}

\nocite{*}
\bibliography{pop_piee}% Produces the bibliography via BibTeX.

\end{document}